\documentclass[pra, aps, nopacs, twocolumn, longbibliography, floatfix,superscriptaddress,amsmath]{revtex4-1}
\usepackage{graphicx}
\usepackage{graphics}
\usepackage{braket}
\usepackage{mathtools}
\usepackage[titletoc]{appendix}
\usepackage{appendix}
\usepackage{float}
\usepackage{bm}
\usepackage{amssymb}

\usepackage[dvipsnames]{xcolor}
\definecolor{darkblue}{rgb}{0.0, 0.0, 0.75}

\usepackage{color}
\usepackage[colorlinks=true,
            linkcolor=darkblue,
            urlcolor=darkblue,
            citecolor=darkblue]{hyperref}

	\usepackage[normalem]{ulem} 
	\definecolor{mgreen}{RGB}{1,123,0}

\def \br{{\bf r}}

\def \mum{\mu \mathrm{m}}

\def \m2D{\mathrm{2D}}

\def \mms{\mathrm{ms}}

\def \mcir{\mathrm{cir}}

\def \mb{\mathrm{b}}
\def \mw{\mathrm{w}}

\def \mr{\mathrm{ring}}

\def \mum{\mu \mathrm{m}}

\def \mkHz{\mathrm{kHz}}

\def \br{\mathbf{r} }

\newcommand*\diff{\mathop{}\!\mathrm{d}}

\begin{document}
\title{Implementation of an  atomtronic SQUID in a strongly confined toroidal condensate}
\author{Hannes Kiehn}
\affiliation{Zentrum f\"ur Optische Quantentechnologien and Institut f\"ur Laserphysik, Universit\"at Hamburg, 22761 Hamburg, Germany}
\author{Vijay Pal Singh}
\affiliation{Zentrum f\"ur Optische Quantentechnologien and Institut f\"ur Laserphysik, Universit\"at Hamburg, 22761 Hamburg, Germany}
\affiliation{Institut f\"ur Theoretische Physik, Leibniz Universit\"at Hannover, Appelstra{\ss}e 2, 30167 Hannover, Germany}
\author{Ludwig Mathey}
\affiliation{Zentrum f\"ur Optische Quantentechnologien and Institut f\"ur Laserphysik, Universit\"at Hamburg, 22761 Hamburg, Germany}
\affiliation{The Hamburg Centre for Ultrafast Imaging, Luruper Chaussee 149, Hamburg 22761, Germany}
\date{\today}
%
%
\begin{abstract}
We investigate the dynamics of an atomtronic SQUID created by two mobile barriers, moving at two different, constant velocities in a quasi-1D toroidal condensate.
We implement a multi-band truncated Wigner approximation numerically, to demonstrate the functionality of a SQUID reflected in the oscillatory voltage-flux dependence.
The relative velocity of the two barriers results in a chemical potential imbalance analogous to a voltage in an electronic system. 
The average velocity of the two barriers corresponds to a rotation of the condensate, analogous to a magnetic flux.
We demonstrate that the voltage equivalent shows characteristic flux-dependent oscillations.
We point out the parameter regime of barrier heights and relaxation times for the phase slip dynamics, resulting in a realistic protocol for atomtronic SQUID operation. 
\end{abstract}

\maketitle
%
%

\section{Introduction}

     Ultracold atom systems provide an ideal platform to study superfluidity and quantum many-body phenomena in well-defined, clean geometries. 
A major breakthrough for the study of superfluidity was the creation of persistent currents of superflow in toroidal Bose-Einstein condensates (BECs) \cite{WrightToroidal, Amico2005, RyuToroidal, RamanathanToroidal,Corman2014}, 
which are analogous to supercurrents in superconducting materials. 
This contributed to the advancement of the field of atomtronics, which aims at the development of atom-based devices \cite{AtomtronicsFirstPaper, Amico2014, Amico2017, Amico2021}, 
e.g. atom-based circuits \cite{Micheli2004, Stickney2007, Ruschhaupt2007, Thorn2008, Pepino2009, Krinner2017, Zozulya2013, Caliga2017, Caliga2016} 
or atom interferometers \cite{Pandey2021, Krzyzanowska2022}. 
Studies on superflow in toroidal BECs have been reported in \cite{Wright2013, Mathey2014, Yakimenko2015, Kumar2016, Kumar2017, Wang2015, Kunimi2019, Polo2019,Zhang2019, Syafwan2016, Bland2020, Mehdi2021}.

  An important and compelling phenomenon is the quantum interference of currents that flow through two Josephson junctions. 
When they are brought to interference the resulting current gives access to the accumulated relative phase. The relative phase is due to the magnetic flux enclosed by the currents. These devices couple to the magnetic field via the electric charge of the Cooper pairs.
This was first observed in superconducting  quantum interference devices (SQUIDs) \cite{clarke2006squid}. 
SQUIDs have applications in quantum sensing \cite{Degen2017} and information processing \cite{Ladd2010}.

The field of atomtronics aims to implement analogues of electronic devices such as SQUIDs. 
Superflow in toroidal atom condensates allows us to realize atom analogs of dc- and rf-SQUIDS using a weak link 
\cite{Wright2013, Eckel2014}, and dc-SQUIDs using two weak links \cite{Ryu2013, Campbell2014}. 
Recently, an atomic dc-SQUID was realized in a voltage state at a constant bias current \cite{Ryu2020}, demonstrating  the quantum interference of currents.

In this paper we propose an experimental setup to measure the voltage-flux dependence of a dc-SQUID, 
which is created by two mobile barriers in a quasi-1D ring condensate. 
We study the system dynamics using a multi-band simulation approach suitable for toroidal condensates in quasi-1D setting. 
Within a truncated Wigner approximation, we take quantum fluctuations into account and extend the theoretical description beyond the Gross-Pitaevskii equation \cite{Ryu2020}.
We simulate the presence of a magnetic field by letting the barriers rotate around the ring. 
This motion excites phase slips in the condensate, similar to how a penetrating flux induces a phase winding in a superconducting loop. 
We show that the average phase winding depends on the barrier height and results in a step-like flux dependence for large barrier heights. 
We then superimpose another barrier motion which lets the barriers approach each other with a nonzero relative velocity. 
This simulates a bias current flowing through the SQUID. 
Operating the SQUID with a bias current results in the formation of a density imbalance, which is analogous to a voltage in an electronic circuit and is a distinct feature of dc-SQUIDs in the resistive regime. 
We analyze the dependence of the voltage on the applied magnetic flux and find characteristic oscillations at the periodicity of the flux quantum. 
This demonstrates quantum interference effects and is analogous to the voltage-flux relation based on electronic SQUID models.  
Our work builds on a previous study on an atomtronic SQUID in a 3D toroidal setting \cite{Mathey2016} and experimental work utilizing a mobile barrier based SQUID without magnetic flux \cite{Campbell2014}.

   This paper is organized as follows. 
In Sec. \ref{section:simulationMethod} we describe the multi-band simulation approach to simulate the dynamics of a quasi-1D condensate. 
In Sec. \ref{section:SQUIDImplementation} we implement the barrier protocol and discuss the operation of an atomtronic dc-SQUID. 
In Sec. \ref{section:phaseSlip} we discuss the emergence of phase-slips due to the barrier motion. 
In Sec. \ref{section:voltageFlux} we analyze the voltage-flux oscillations.  
We conclude in Sec. \ref{sec:con}.

\section{Simulation method}\label{section:simulationMethod}

  We consider a lithium gas in a toroidal trap, in which the lithium atoms are in molecular form of two bound $^6$Li atoms. The trapping potential is harmonic along the transverse directions with the trapping frequencies $\omega_{\rho}$ and $\omega_z$, see below. 
The trapping energies are larger than the mean-field energy $\mu$ of the gas, i.e. $\hbar \omega_{\rho, z} > \mu$. Therefore, the system is in a quasi-1D regime. We emphasize that we do not approximate the gas as a purely-1D system, via a single-band approximation, but rather include transverse motion by including several of the lowest bands, resulting in a quasi-1D description, as we expand on below. 
Spatially, the condensate is confined to several $\mum$ in the transverse directions.
The underlying Hamiltonian of a 3D interacting BEC is
\begin{equation}
 \hat{H}=\int \diff{\br}\left( \hat{\psi}^\dagger\left[-\frac{\hbar^2 \nabla^2}{2m_D}+V \right] \hat{\psi}+\frac{g}{2}\hat{\psi}^\dagger\hat{\psi}^\dagger \hat{\psi}\hat{\psi}\right),
\end{equation}
where $g=4\pi a_s \hbar^2/m_D$ is the interaction strength proportional to the s-wave scattering length $a_s$, and $m_D$ is the molecular mass, i.e. $m_D = 2 m_{\mathrm{Li}}$. 
The potential $V(\br)$ is a harmonic ring trapping potential $V_{\text{trap}} (\br) = m_D \bigl[ \omega_\rho^2 (\rho - R)^2 + \omega_z^2 z^2 \bigr]/2$.
We set the ring circumference to $250\, \mum$, so the radius is $R_\mr \approx 40\, \mum$. 
The trapping frequencies are $\omega_\rho = 2 \pi \times 4 \, \mkHz$ and $\omega_z = 2 \pi \times 1.5 \, \mkHz$.
For our numerical calculation, we discretize the spatial motion along the azimuthal direction  on a 1D lattice of 250 sites with discretization length  $l=1\, \mum$  \cite{Mora2003}. 
We work in cylindrical coordinates $(\rho,\theta,z)$. 
The transversal degrees of freedom $\rho$ and $z$ are treated by expanding the field operator $ \hat{\psi}$ for each site $i$ as
\begin{equation}
 \hat{\psi}^\dagger_i=\sum_{m_1,m_2} \phi_{m_1}(\rho)  \phi_{m_2}(z)  \hat{\psi}_{m,i}^\dagger,
\end{equation}
where $\phi$ are the eigenfunctions of the quantum harmonic oscillator. 
The quantum numbers $m=(m_1, m_2)$ are written as a tuple for compact notation.
We restrict ourselves to $0\leq m_1,m_2 \leq 2$, i.e. we include two excited states in each transverse direction. With this expansion in higher bands in the transverse directions, the system is modelled as a multi-band Hubbard model. 
For our numerical simulations we use the truncated Wigner approximation and therefore approximate the quantum dynamics by a semiclassical evolution \cite{Blakie2008, Polkovnikov2010}. This allows us to replace the operators $ \hat{\psi}$ by complex numbers $\psi$. We sample the initial state from a continuous Wigner function, which has the form of a Gaussian for a non-interacting system \cite{Polkovnikov2010}. 
Each state of the initial ensemble propagates under the classical equations of motion
\begin{align}
  i \hbar \frac{\diff{\psi_{k,i}}}{\diff{t}} &= \hbar \bigl( \omega_\rho (m_1+1/2)+\omega_z(m_2+1/2) \bigr)\psi_{k,i}    \nonumber  \\
  &-J\left(\psi_{k,i+1}+\psi_{k,i-1}-2\psi_{k,i} \right)    \nonumber  \\ 
  &+\sum_{lmn} \psi_{l,i}^* \psi_{m,i} \psi_{n,i} U_{klmn}+ \sum_{n} V_{\text{b},k,n,i} \psi_{n,i}, 
\end{align}
where  $J=\hbar^2/(2m_D l^2)$ is the tunneling energy and $V_{\text{b},k,n,i}$ correspond to the external barrier potential $V_{\text{b}}$ that we introduce in Sec. \ref{section:SQUIDImplementation}. 
The coefficients for the onsite interaction and the barrier potential  are defined as
\begin{equation}
 U_{klmn}=\frac{g}{l}\int \diff{\rho}\phi_{k_1}\phi_{l_1}\phi_{m_1}\phi_{n_1} \int \diff{z}\phi_{k_2}\phi_{l_2}\phi_{m_2}\phi_{n_2}
\end{equation}
and 
\begin{equation}
V_{\text{b},k,n,i} = \int \diff{\rho}\diff{z} V_{\text{b},i}\phi_{k_1}  \phi_{k_2} \phi_{n_1} \phi_{n_2},
\end{equation}
where the basis functions $\phi_j$ are taken to be real. 
These coefficients are given in Appendix \ref{app:coefficients}.

   As described above, we initialize the dynamics by sampling the Wigner function of a non-interacting system at zero temperature. The ground state is populated by a large number of particles $N=40000$. 
All excited states only contain quantum fluctuations leading formally on average to an increase of $1/2$ particles per state or lattice site. The interaction is then slowly increased to the desired value of $U/J = 0.01$ over $3.8$ s. 
This results in heating of the condensate to a temperature comparable to the mean-field energy.
The resulting state is the initial state of the simulation of the physical system, for which we set the initial time to $t=0$. 
We add two mobile barriers to create an atomtronic SQUID, as we describe in Sec. \ref{section:SQUIDImplementation}. 
As our observables, we calculate the local density $n(\theta) = |\psi(\theta)|^2$ and the global phase winding $\Omega_\mw = \sum_\theta \delta \phi_0 (\theta )$, where the phase difference $ \delta \phi_0 (\theta ) = \phi_0 (\theta + l) - \phi_0 (\theta) $ is between $-\pi$ and $\pi$. 
$\phi_0 (\theta) $ is the local phase corresponding to the lowest (condensate) mode.
We calculate these observables for each sample and then average over the initial ensemble to obtain the average density and the average phase winding.

\begin{figure}[]
\centering
\includegraphics[width=0.45\textwidth]{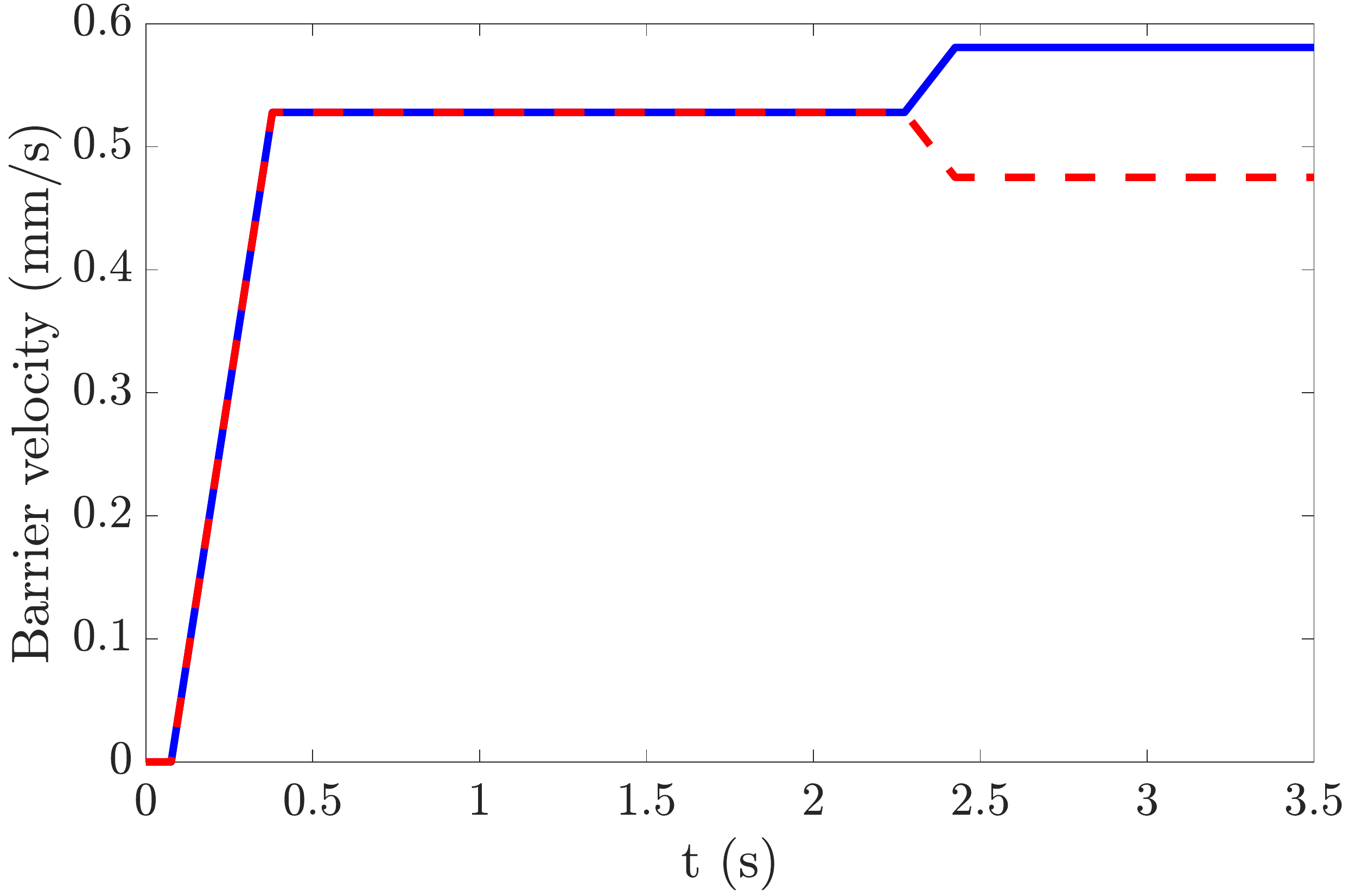}
\caption{\label{fig:fig0} Barrier protocol. 
We initially turn on the barrier strength $V_0$ over $75\, \mms$ and then slowly ramp up the barrier velocities $v_{1,2}=530 \, \mu \mathrm{m/s}$ over  $300\, \mms$. 
The barrier-induced stirring motion sets the condensate in rotation, which relaxes to the phase winding state of lowest energy over a stirring time of $2.2$ s.  
After that, we ramp up the bias velocity  $v_\text{bias}=106 \, \mu \mathrm{m/s}$ over $150\, \mms$, 
resulting in different $v_{1}$ and $v_{2}$.  
}
\end{figure}
\begin{figure*}[ht]
\centering
\includegraphics[width=0.825\textwidth]{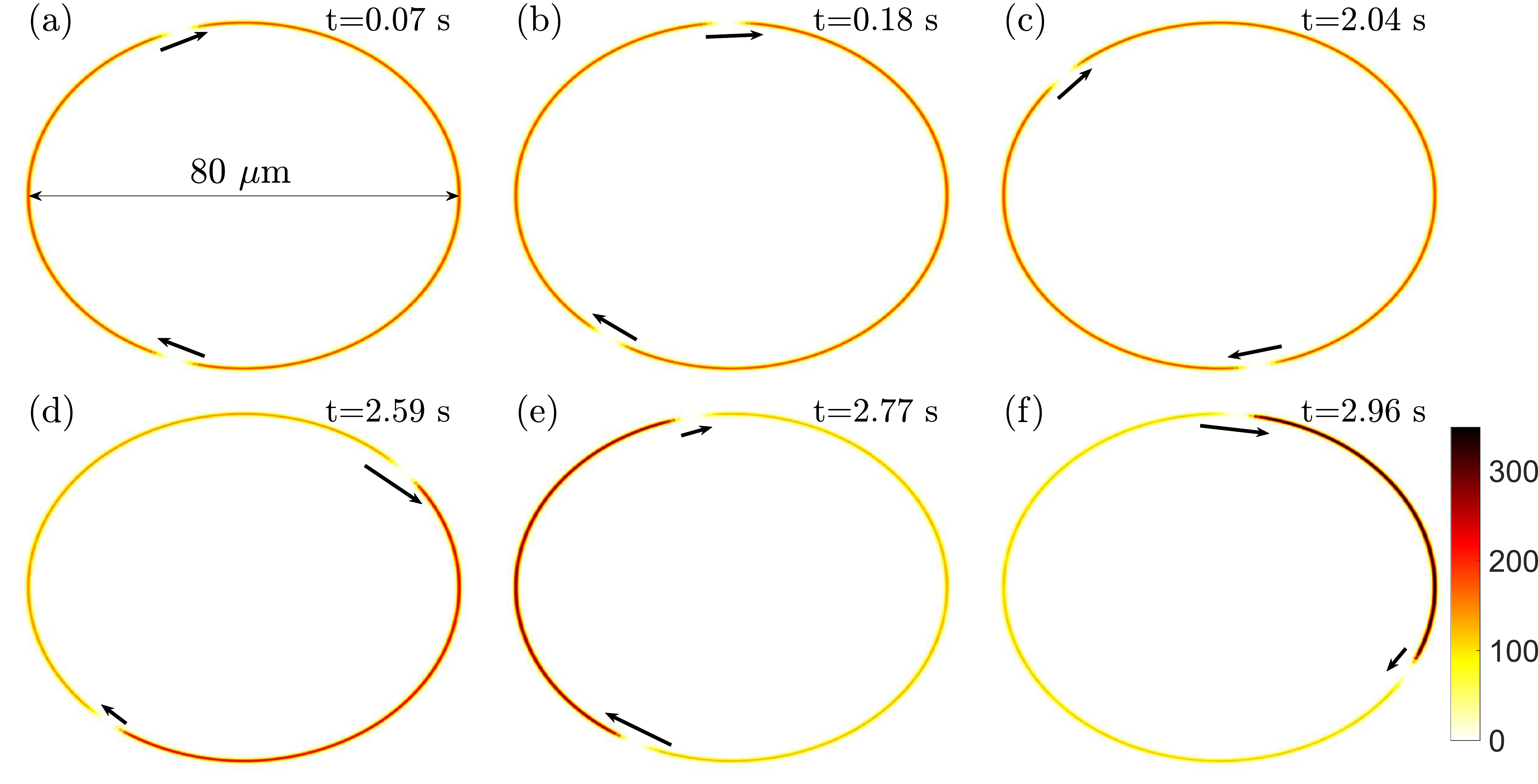}
\caption{SQUID dynamics in a ring condensate. Column density $n(\rho, \theta) =|\psi (\rho, \theta)|^2$ at different times during the barrier protocol with the barrier strength $V_0/\mu=1.7$. 
At earlier times (a-c), the two barriers move with the same velocity $v_\phi$ by choosing $v_{1,2}= 530\, \mu  \mathrm{m/s}$ and $v_\text{bias}=0$, which models a magnetically induced screening current.  
The arrows indicate the barrier velocities and direction of motion.
The  stirring motion continues for $2.2$ s, which allows the condensate to relax in the phase winding state of the lowest energy. 
At later times times (d-f), the barriers move at different velocities $v_{1,2}=\pm v_{\text{bias}}/2 + v_{\Phi}$, with a bias velocity of  $v_\text{bias}=106 \, \mu \mathrm{m/s}$, and thereby start to approach each other. 
The strong barrier results in a resistive regime and a density imbalance is formed, in analogy to a voltage.}
\label{fig:fig1}
\end{figure*}
%


%
\section{Implementation of an atomtronic SQUID}\label{section:SQUIDImplementation}
A solid state realization of a dc-SQUID consists of two Josephson junctions in a superconducting ring. 
In the following, we implement such a device as an atomtronic system in the so-called voltage state, where a constant bias current $I_{\text{bias}}$ is applied to the setup. Furthermore, we apply the equivalent of a magnetic flux $\Phi$ through the ring. 
As we demonstrate below, the voltage shows characteristic oscillations depending on the magnetic flux that is applied to the SQUID, in analogy to a solid state SQUID.
The required current regime derives from the critical current $I_c$ of a Josephson junction, at which the junction switches from the non-resistive to the resistive regime. 
Given the two Josephson junctions in the SQUID, the minimal bias current is $2 I_c$, for a nonzero voltage across the SQUID.

 The existence of these voltage-flux oscillations is due to the fact that the flux can only pass through the ring in quantized multiples of the flux quantum $\Phi_0$. 
If the magnetic flux is not an integer multiple of $\Phi_0$, i.e.  $\Phi\neq n \Phi_0$  $(n=0,1,2,..)$, 
a screening current $I_s$ is formed in the ring to enhance or reduce the flux to an integer multiple of $\Phi_0$. 
Therefore, the total current at the two junctions is modified to $I_{\text{junction}} = I_{\text{bias}}/2 \pm I_s$. This leads to a lowered critical current for higher $|I_s|$ which in turn increases the voltage. So the voltage-flux curve oscillates with a periodicity of $\Phi_0$ \cite{tinkham2004}.

An atomic analogue of solid-state SQUID is experimentally realized by Ref. \cite{Ryu2020}. 
Here we propose to measure voltage-flux oscillations as follows. 
We use two barrier potentials, of the form:
\begin{equation}
 V_\text{b}(\br)= V_0 e^{-\rho^2} \left[e^{-(\theta_{\mb 1}-\theta)^2/(2\sigma^2)}+e^{- (\theta_{\mb 2}-\theta)^2/(2\sigma^2)}\right]
\end{equation}
to model the Josephson junctions of width $\sigma=0.9\, \mum$. 
$\theta_{\mb 1}$ and $\theta_{\mb 2}$ are the angular position of the two barriers.
Unlike the electronic SQUID where the barriers are stationary, we consider mobile barriers,
 i.e.  $\theta_{\mb 1, \mb 2}=\theta_{\mb 1, \mb 2} (t)$.
Instead of letting a current $I_{\text{bias}}$ flow through the junction, we move the barriers with a velocity $\pm v_{\text{bias}}/2$. Note that this  scenario of equal velocity barriers has been experimentally realized by Ref. \cite{Campbell2014}. 
To model the magnetic flux we add the velocity $v_{\Phi}$, so that the two velocities are $v_{1,2}=\pm v_{\text{bias}}/2 + v_{\Phi}$.

The barrier protocol is described in Fig. \ref{fig:fig0}.  
We gradually ramp up the barrier strength $V_0$ over $75$ ms. 
Inspired by the setup in Ref. \cite{Campbell2014} we set the initial distance between the barriers to $x_0=100 \, \mum$, i.e.  a value less than half the circumference. This provides more space for the barrier movement later on. Then we proceed to slowly accelerate both barriers to $v_\Phi$ by choosing the same $v_{1,2}$ and $v_\text{bias}=0$, as depicted in Fig. \ref{fig:fig0}.
At this stage the barriers move in the same direction with equal velocity, which we show in the simulations in Figs. \ref{fig:fig1}(a-c). 
The distance between the barriers remains constant. Due to this stirring motion, a phase winding is formed in the condensate, indicating rotary motion. 
As we  elaborate on in the next section, the relaxation of the condensate to the phase winding state of lowest energy, without suppressing the phase coherence entirely, is slow for typical experimental time scales.
Therefore, we continue the stirring motion for $2.2$ s for the system to relax. 
We note that in a typical solid state SQUID, the relaxation time to the current state with the lowest energy is fast compared to the time scales of the operation. In an atomtronic SQUID these time scales are in general not strongly separated, which provides an intrinsic challenge of atomtronics to emulate solid state devices.
 
  The final step is to slowly increase $v_{\text{bias}}$ to its final value over the course of $150$ ms,  as indicated in Fig. \ref{fig:fig0}.  
 At this point the barriers move with unequal velocities $v=\pm v_{\text{bias}}/2 + v_{\Phi}$ as can be seen in the simulation results 
in Figs. \ref{fig:fig1}(d-f). We show the example $v_{\text{bias}}/2 < v_{\Phi}$ so that both barriers still move in the same direction, but with different velocities. 
As a result, a density difference between the two subsystems can emerge, depending on the regime of operation.
Within the superfluid regime of the Josephson junctions, the atoms tunnel through the barriers and the densities are equal in the two subsystems. 
However, in the resistive regime the tunnel current is not sufficiently large to maintain a density balance. Hence, the condensate density increases on the side of the barriers on which the two barriers approach each other and  
a density difference emerges in the system [Fig. \ref{fig:fig1}(f)], analogous to a voltage in an electronic SQUID.

\section{Phase slip dynamics}\label{section:phaseSlip}
\begin{figure}[t]
\centering
\includegraphics[width=0.5\textwidth]{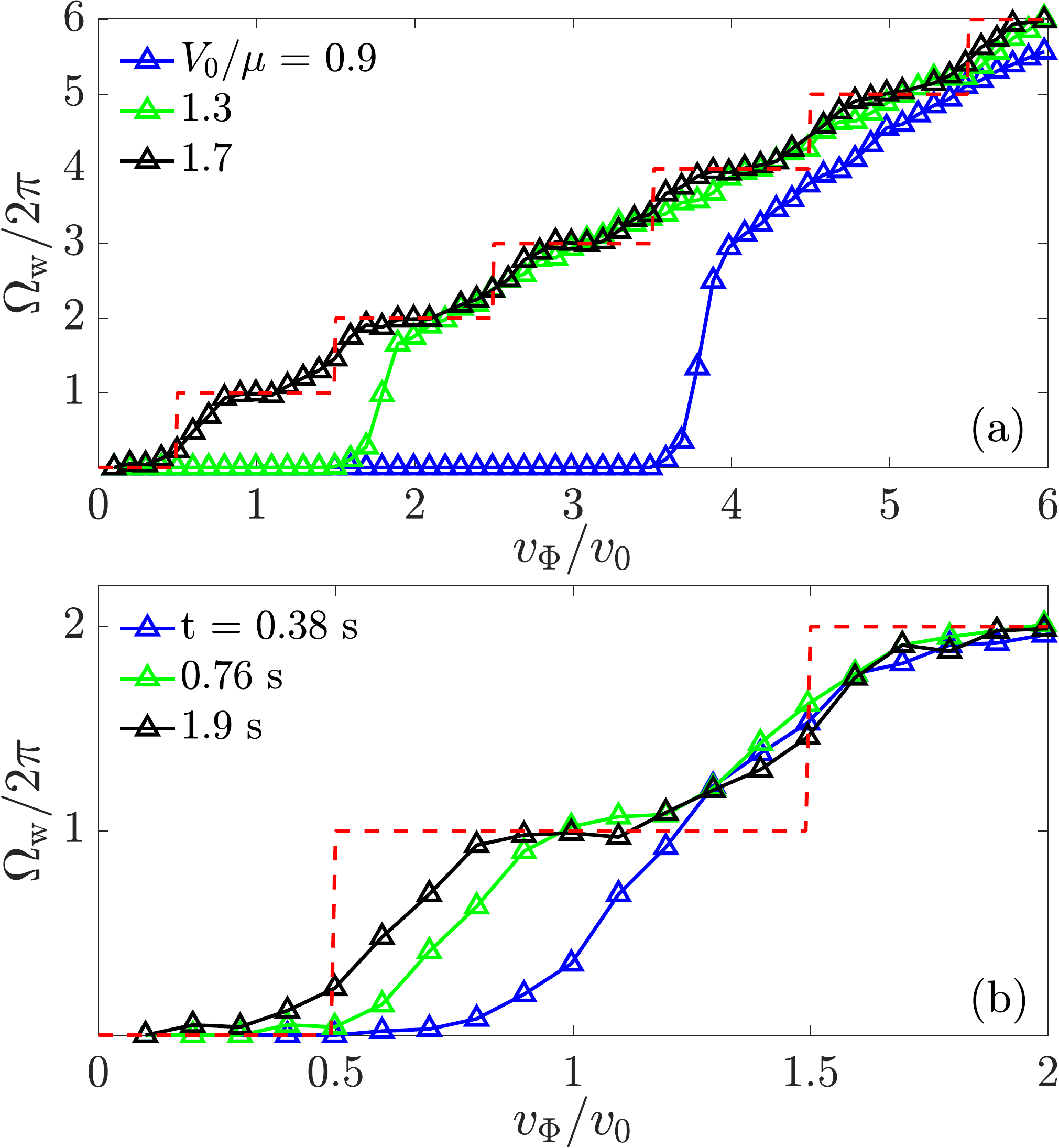}
\caption{Phase-slip dynamics (a) Average phase winding $\Omega_\mw$ as a function of the flux velocity $v_\Phi/v_0$ at time $t = 1.9$ s for the barrier height of $V_0/\mu =0.9$, $1.3$ and $1.7$. 
For  $V_0/\mu =1.7$ the system dynamics approaches the idealized step-like behavior (dashed line). 
(b)  $\Omega_\mw(t)$ for $V_0/\mu =1.7$ at  $t = 0.38$, $0.76$ and $1.9$ s. 
The system needs considerable time,  for typical experimental operation, for the phase winding to relax into a step-like pattern.}
\label{fig:fig2}
\end{figure}

In an electronic SQUID, the magnetic flux penetrates the ring in integer multiples of $n \Phi_0$, with $n$ being an integer. 
The wave function in the ring acquires a phase winding of $2 \pi n$ accordingly. 
As the SQUID relaxes to its state of lowest energy, 
phase slips occur at $\Phi=(n+1/2)\Phi_0$, because this minimises the flux provided by the screening current to $\leq \Phi_0/2$.
We demonstrate the analogous behaviour for the atomtronic SQUID.
Here, the relative velocity between condensate and barriers is minimised by the rotation of the condensate. 
Every $2\pi$ of phase winding is associated with a rotation velocity $v_0=2\pi\hbar/(m_D R_\mr)$. 
This velocity $v_0$ plays the role of a flux quantum in this system. 
So the phase slips occur at $v_\Phi=(n+1/2)v_0$, if the condensate relaxes to the lowest energy state.
We investigate this phase slip dynamics for $v_{\text{bias}}=0$ and nonzero $v_\Phi$ as in Figs. \ref{fig:fig1}(a-c).
We calculate the average phase winding $\Omega_\mw$ as described in Sec. \ref{section:simulationMethod} and analyze its dependence on the stirring time $t$ and  barrier height $V_0$.  
In Fig. \ref{fig:fig2}(a) we show the average phase winding in the system at $t=1.9$ s,  for different values of $V_0$.
As a dashed line, we show the phase winding increasing in unit steps, at half-integers of the flux quantum, as described above.
The average phase winding that has accumulated in the condensate at that time deviates from the idealized, equilibrium expectation (dashed line). 
For low barriers ($V_0/\mu =0.9$), no phase slips are observed for flux velocities up to $v_\Phi / v_0\sim 3.5$.  
For $v_\Phi / v_0 \gtrsim 4$, the dependence of $\Omega_\mw$ on $v_\Phi$ is approximately linear, rather than an approximate step-like behavior. 
So the condensate does not  relax to the rotational state of lowest energy, and SQUID operation is not possible. 
The origin of the slow relaxation, and the condensate remaining in a metastable state, is that the critical velocity at the barrier is too high. 
To emulate an electronic SQUID, the condensate has to relax to the lowest energy phase winding, to  be in a resistive state for phase slips to occur. 
For that, we modify the barrier height.  
Indeed for $V_0/\mu=1.3$ the critical velocity is lower and the steps are more pronounced while not sufficiently so to be a SQUID. 
A good step-like behaviour is observed for $V_0/\mu=1.7$, 
where the critical velocity is reduced to a value similar to $v_\Phi / v_0\sim 0.5$.
Note that the relaxation is generally more effective near the center of a step, i.e. at whole integer values, where screening currents are relatively low. At half integer values there are two ground states, one with $v_s = v_0/2$ and another one with $v_s = -v_0/2$ plus a phase winding. These states have nearly equal energy at this rotation velocity, and the tendency of the system to relax is suppressed.
We note that such barrier-height induced deviations for phase slips were also studied in a single-barrier based atomic SQUID \cite{Mathey2016}  and the suppression of the phonon velocity due to the barrier height was pointed out in toroidal condensates \cite{Mathey2014}, which is similar to the reduction of the critical velocities in stirred BECs \cite{Singh2016, Kiehn2021}.

 To illustrate the relaxation process, we consider the optimal barrier height $V_0/\mu=1.7$ and analyze the time evolution of 
 $\Omega_\mw(t)$ at different stirring times. 
The results are shown in Fig. \ref{fig:fig2}(b), where we focus on the first phase slip at $v_\Phi / v_0 = 0.5$.  
The behaviour for different barrier heights is similar for larger $v_\Phi$ as we saw in Fig. \ref{fig:fig2}(a). 
At $t=0.38$ s the system is not relaxed as the step-like behaviour is not visible. 
At $t=0.76$ s, we see the emergence of a clear plateau in the phase. 
Finally, at $t=1.9$ s  the phase is reasonably converged and a clear plateau is visible at $2\pi$.

\section{Voltage-flux relation}\label{section:voltageFlux}

\begin{figure}[t]
\centering
\includegraphics[width=0.5
\textwidth]{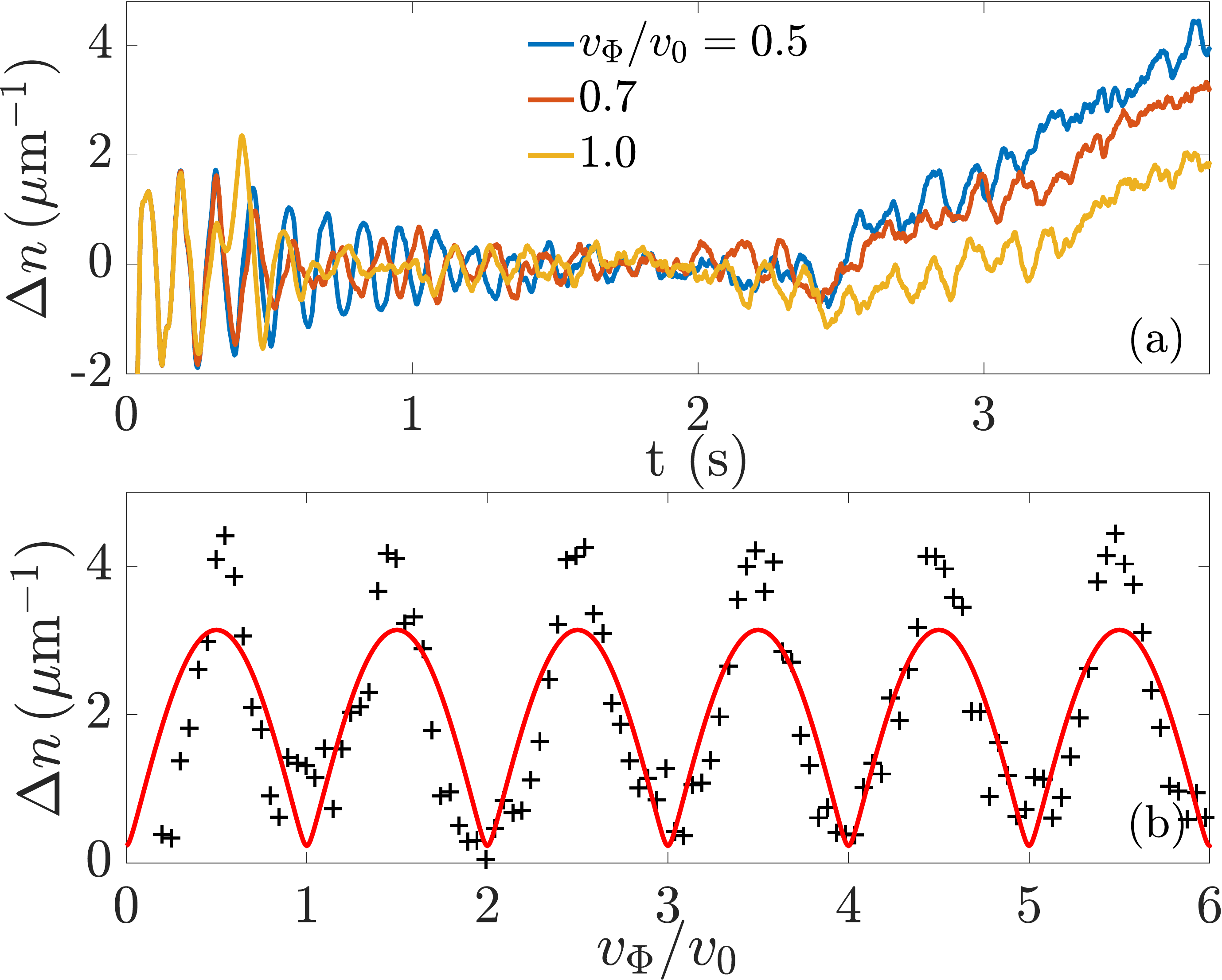}
\caption{Voltage-flux oscillations. (a) Time evolution of the density imbalance $\Delta n(t)$ at the flux velocity of $v_\Phi/v_0=0.5$, $0.7$ and $1$. 
A strong imbalance is observed at $v_\Phi/v_0=0.5$, as contrary to the result at $v_\Phi/v_0=1$.
(b) Final density imbalance $\Delta n$ at $t=3.8\, \mathrm{s}$ as a function of $v_\Phi/v_0$. 
The continuous line is the fit with the flux-voltage relation of a SQUID in Eq. \ref{eq:fit}.}
\label{fig:fig3}
\end{figure}

As described earlier, we propose to induce a density imbalance in the ring by the moving barriers. 
This density imbalance, which corresponds to a chemical potential difference, is analogous to the voltage that emerges in a SQUID within the resistive regime.
We now discuss this imbalance in more detail and its dependence on the rotational flux for $V_0/\mu =1.7$.
We determine the density imbalance $\Delta n = n_R - n_L$, where $n_R$ and $n_L$ are defined as the density of the initially larger and smaller segment of the ring, respectively, see Sec. \ref{section:simulationMethod} and Fig. \ref{fig:fig1}(a). 
In Fig. \ref{fig:fig3}(a) we show the time evolution of $\Delta n (t)$ at $v_\Phi/v_0 =0.5$, $0.7$ and $1$. 
Early in the dynamics, small density oscillations are exited by the motion of the barrier. 
These are the equivalent of plasma oscillations \cite{Levy2007, Albiez2005} and create some undesirable noise in the  system. We note that these oscillations are observed experimentally as well \cite{Campbell2014}. 
However, with a relative amplitude of $\sim 2.5\, \%$ they are more significant than the ones observed in the experiment, where $\Delta n / n \sim 1\, \%$. 
Over 2 s they damp out significantly, and to fluctuations with relative amplitudes of $\Delta n / n \sim 1\, \%$.

 In the absence of a bias current,  the average imbalance of the condensate is zero, as shown in the time evolution before $t \sim 2.4$ s in Fig. \ref{fig:fig3}(a). After the relaxation process at $t \sim 2.4$ s, we slowly turn the bias velocity to a maximum value of $v_{\text{bias}}=10.6\, \mum/\mathrm{s}$, which is a factor of $10$ smaller than the value used in Fig. \ref{fig:fig1}. 
This results in a linear growth for $\Delta n (t)$, visible in Fig. \ref{fig:fig3}(a). 
The magnitude of  $\Delta n (t)$ depends on the flux velocity $v_\Phi/v_0$.
For half integer multiples, i.e. $v_\Phi/v_0 = (n+1/2)$, we expect a high resistance, i.e. strong imbalance, because the critical velocity is minimal at this point.
At whole integer values the resistance should be minimal as no screening currents are present in the system. 
This is consistent with the results shown in Fig. \ref{fig:fig3}(a). 
We repeat the above protocol for multiple values of $v_\Phi/v_0$ and extract $\Delta n (t)$ at $t=3.8$ s using a linear fit. 
We show these results in Fig. \ref{fig:fig3}(b). As described above, we find a minimum of $\Delta n$ at full integer values of $v_\Phi/v_0$ and a maximum at half integer values of $v_\Phi/v_0$. 
This dependence on $v_\Phi/v_0$ results in oscillations of the density imbalance, which is analogous to voltage-flux oscillations of a solid-state SQUID. 
The observed oscillations are direct evidence for quantum interference of currents.
The voltage-flux relation of the dc-SQUID in the overdamped limit is \cite{clarke2006squid}
\begin{equation}\label{eq:fit}
 \braket{V(t)} = I_c R \sqrt{ \frac{I}{2 I_c} - \left(\cos(\pi \frac{\Phi}{\Phi_0}) \right)^2 },
\end{equation}
which assumes negligible screening and yields the time averaged junction voltage as a function of the applied flux $\Phi$. 
$I_c$ is the critical current of a single junction, $I$ is the total current, $R$ is the resistance and $\Phi_0$ is the magnetic flux quantum. 
We emphasize that the condensate realization of a SQUID is not in the overdamped regime, and that we merely use this expression as a fitting function, and more generally as a comparison, see also Appendix \ref{app:damp} for comparison to a dc-SQUID model with non-zero capacitance and inductance.
We fit this expression to our results and show the fit as a continuous line in Fig. \ref{fig:fig3}. 
We find a qualitatively good agreement with the analytic expression of Eq. \ref{eq:fit}. 
In particular, the periodic nature of the dependence is captured, which is the central, defining property of a SQUID.
However, the fit seems to underestimate the amplitude of the oscillations,  which is captured better by the nonzero capacitance and inductance dc-SQUID model as described in Appendix \ref{app:damp}.
From the fit, we extract the parameters $I_c R = 3.1$ and $I/(2 I_c)= 1.005$. 
The current is more than twice the critical current, which is consistent with the SQUID being in the resistive regime.

\section{Conclusions} \label{sec:con}

In conclusion, we have put forth a proposal of how to create an atomtronic SQUID in a toroidal quasi-1D condensate. For that purpose, we simulate the condensate dynamics via a numerical implementation of a multi-band truncated Wigner approximation. We note that this method is ideally suited for the dynamics of fluctuating quasi-1D condensates, such as in thin-ring geometries \cite{Pandey2019}.
We stir the condensate with two mobile barriers to induce rotation of the cloud, which is equivalent to a magnetic flux applied to a conventional SQUID.
This allows us to study the phase-slip dynamics and  its dependence on the stirring time and the barrier height. 
For long stirring times and strong barrier heights the average phase winding results in a step-like behavior as a function of the flux velocity.  We then operate the SQUID in this regime at a constant bias velocity to create a density imbalance, which shows characteristic oscillations with a periodicity of the flux velocity quantum. 
This highlights the voltage regime for the operation of atomtronic SQUIDs and demonstrates the quantum interference of currents.

\section{Acknowledgements}
We thank Kevin Wright, Luigi Amico and Giacomo Roati for insightful discussions.  
This work is supported by the Deutsche Forschungsgemeinschaft (DFG) in the framework of SFB 925 – project ID 170620586 
and the excellence cluster  `Advanced Imaging of Matter’ - EXC 2056 - project ID 390715994, and
 the Cluster of Excellence ‘QuantumFrontiers’ - EXC 2123 - project ID 390837967.

%
%
\appendix
\setcounter{equation}{0}
\setcounter{figure}{0}
\setcounter{table}{0}
\setcounter{section}{0}
\setcounter{subsection}{0}
\renewcommand{\thefigure}{A\arabic{figure}}
\renewcommand{\thetable}{A\arabic{table}}

\section{Numerical Coefficients}\label{app:coefficients}
We define dimensionless coefficients $c$ via
\begin{equation}
  U_{klmn} = \frac{U}{2 \pi l_{\text{osc},\rho} l_{\text{osc},z} } c_{k_1l_1m_1n_1} c_{k_2l_2m_2n_2},
\end{equation}
where $U=g/l^3$ and $l_{\text{osc}}$ is in units of the discretization length $l$. These are given by an integral
\begin{equation}
  c_{k_1l_1m_1n_1}=\sqrt{2 \pi}\int \diff{x}\phi_{k_1}\phi_{l_1}\phi_{m_1}\phi_{n_1},
\end{equation}
where
\begin{equation}
 \phi_{n}(x)=(\pi)^{-1/4} (2^n n!)^{-1/2}\exp(-\frac{x^2}{2})H_n(x)
\end{equation}
are dimensionless harmonic oscillator functions. The normalization of of the $c$ coefficients is chosen such that $c_{0000}=1$. Note that due to the symmetry properties of the Hermite functions, $c_{abcd}$ is zero when $a+b+c+d$ is odd. 
Also $c_{abcd}$ is invariant under permutation of the indices. We provide the coefficients used in our calculation in table.

\begin{table}[h]
\caption{Values of interaction coefficients. }
\begin{tabular}{ |p{1cm}||p{1.0cm}|  }
 \hline
 \multicolumn{2}{|c|}{Coefficients} \\
 \hline
 $c_{0000}$& 1\\
 $c_{1100}$& $\frac{1}{2}$  \\
 $c_{1111}$& $\frac{3}{4}$\\
 $c_{2000}$& $-\frac{1}{2 \sqrt{2}}$\\
 $c_{2110}$& $\frac{1}{4 \sqrt{2}}$\\
 $c_{2200}$& $\frac{3}{8}$\\
 $c_{2211}$& $\frac{7}{16}$\\
 $c_{2220}$& $\frac{1}{16 \sqrt{2}}$  \\
 $c_{2222}$& $\frac{41}{64}$\\ [1ex] 
 \hline
\end{tabular}
\end{table}

The coefficients for the barrier potential are determined by
\begin{align}
  V_{b,k,n,i} &= \int \diff{\rho} \diff{z} V_{b,i} \phi_{k_1} \phi_{k_2} \phi_{n_1} \phi_{n_2}\\
   &= \int \diff{\rho} V_{b,i} \phi_{k_1} \phi_{n_1} \delta_{k_2,n_2}.
\end{align}
This results in
\begin{align}\label{eq:bcoeff}
 V_{b,k,n,i}   &= V_0 \delta_{k_2,n_2} \big[e^{-(\theta_{\text{b1}}-\theta)^2/(2\sigma^2)} + e^{-(\theta_{\text{b2}}-\theta)^2/(2\sigma^2)}\big]  \nonumber \\
 & \quad \times \int \diff{\rho}  e^{\rho^2/l_\rho^2} \phi_{k_1} \phi_{n_1}.
\end{align}
We numerically evaluate the integral in Eq. \ref{eq:bcoeff} to obtain the coefficients.

\section{Voltage-flux relation of dc-SQUIDs}\label{app:damp}

For practical SQUIDs the inductance of the circuit and the capacitance of Josephson junctions are taken into account, corresponding to the circulating screening current and the junction displacement current, respectively.  This general case of the dc-SQUID circuit is described by  the equations \cite{GrossNotes}
\begin{align}
\frac{I}{2} =& \frac{\hbar C}{ 2 e} \frac{d^{2} \phi_{1}}{d t^{2}} + \frac{\hbar}{2 e R_{N}} \frac{d \phi_{1}}{d t} + I_{c} \sin\phi_{1} - I_{\mcir}\\
\frac{I}{2} =& \frac{\hbar C}{ 2 e} \frac{d^{2} \phi_{2}}{d t^{2}} + \frac{\hbar}{2 e R_{N}} \frac{d \phi_{2}}{d t} + I_{c} \sin\phi_{2} + I_{\mcir}  
\end{align}
 with the integer constraint
\begin{align}
 2 \pi n =& \phi_{2} - \phi_{1} - 2 \pi \frac{\Phi}{\Phi_{0}} - 2 \pi \frac{L I_{\mcir}}{\Phi_{0}}
\end{align}
with $\Phi_{0} = \frac{\pi \hbar}{e}$ being the flux quantum and $\Phi$ being the applied flux. $\phi_1$ and  $\phi_2$ are the phase differences across the junctions, $I_c$ is the critical current  and $C$ is the capacitance of single junction. $I_\mcir$ is the circulating screening current and $L$ is the inductance.  $R_N$ is the resistance.

We introduce
\begin{align}
\bar{t} = \frac{R_{N} I_{c}}{\Phi_{0}} t,  \quad \bar{\Phi} = \frac{\Phi}{\Phi_{0}}, \quad  \bar{I} = \frac{I}{I_{c}},  \quad \bar{I}_{\mcir} =& \frac{I_{\mcir}}{I_{c}}
\end{align}
and
\begin{align}
\beta_{L} =& \frac{2 L I_{c}}{\Phi_{0}}\\
\beta_{C}=& \frac{2\pi I_{c} R_{N}^{2} C}{\Phi_{0}}
\end{align}
and
\begin{align}
\phi_{\mcir} =& \pi \beta_{L} I_{\mcir}.
\end{align}
For parameter $\beta_{L}$ we have
\begin{align}
\beta_{L} = \frac{2 m R_N I_{c}}{\hbar n_{1d}}.
\end{align}
$n_{1d}$ is the average density projected onto the spatial degree of freedom along the condensate.

\begin{figure}[t]
\centering
\includegraphics[width=0.5\textwidth]{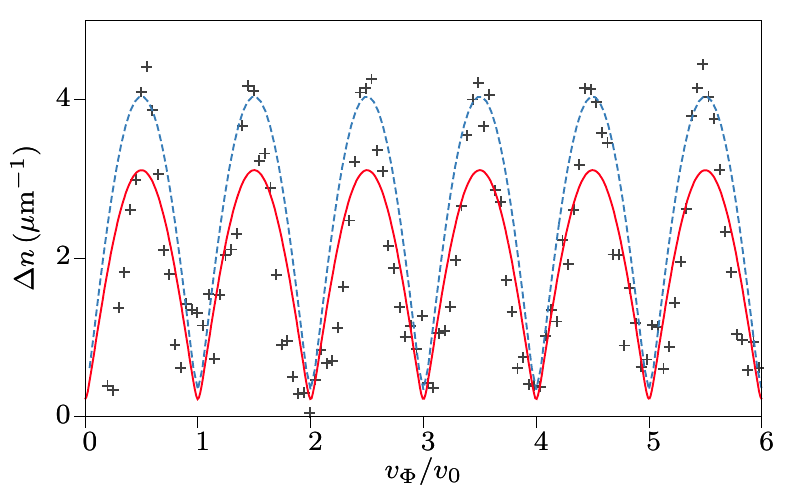}
\caption{Simulated density imbalance $\Delta n$ as a function of $v_\Phi/v_0$, which is the same as Fig. \ref{fig:fig3}(b).
The continuous line is the fit with the over-damped model of dc-SQUIDs in Eq. \ref{eq:fit}, while the dashed line corresponds to the nonzero capacitance and inductance SQUID model of Eq. \ref{eq:vphi}. }
\label{app:fig2}
\end{figure}

  In terms of rescaled units we have:
\begin{align}
\frac{\bar{I}}{2} =& \frac{\beta_{C}}{ 4 \pi^{2}} \frac{d^{2} \phi_{1}}{d \bar{t}^{2}} + \frac{1}{2\pi} \frac{d \phi_{1}}{d \bar{t}} +  \sin\phi_{1} - \bar{I}_{\mcir}\\
\frac{\bar{I}}{2} =& \frac{\beta_{C}}{ 4 \pi^{2}} \frac{d^{2} \phi_{2}}{d \bar{t}^{2}} + \frac{1}{2\pi} \frac{d \phi_{2}}{d \bar{t}} +  \sin\phi_{2} + \bar{I}_{\mcir}
\end{align}
and
\begin{align}
 2 \pi n =& \phi_{2} - \phi_{2} - 2 \pi \bar{\Phi} -  \pi \beta_{L} \bar{I}_{\mcir}.
 \end{align}
 We fulfill this constraint via
\begin{align}
 \bar{I}_{\mcir} =& \frac{2}{\beta_{L}} \Big( \frac{\phi_{2}-\phi_{1}}{2\pi} - \bar{\Phi} - R\Big(\frac{\phi_{2}-\phi_{1}}{2\pi} - \bar{\Phi}\Big) \Big),
\end{align}
 where $R(x)$ rounds the real number $x$ to the nearest integer number.
 The voltage is
\begin{align}\label{eq:vphi}
  V =& \frac{R_{N} I_{c}}{4 \pi} \Big(\frac{d \phi_{1}}{d \bar{t}}  + \frac{d \phi_{2}}{d \bar{t}}  \Big).
\end{align}
We numerically solve Eq. \ref{eq:vphi} to determine the voltage-flux relation $V(\phi)$. 
The screening parameter $\beta_L$ and the Stewart-McCumber parameter $\beta_C$ both have to be smaller than unity to avoid hysteretic $V(\phi)$ curves. As an example, we therefore choose $\beta_C=0.1$ and $\beta_L = 0.1$. 
$\bar{I} =2.01$ is chosen according to the fit result in Sec. \ref{section:voltageFlux}. 
We calculate the voltage-flux relation and scale its magnitude by the factor $\alpha=2.04$, which accounts for the resistance, i.e.,  $\alpha= R_N I_c$. 
This result is shown in Fig. \ref{app:fig2}, which describes the simulation result better than the over-damped model. 
This is also  reflected by the value of the resistance being smaller than that of the over-damped fit. 


%
\bibliography{sample}

\begin{thebibliography}{50}%
\makeatletter
\providecommand \@ifxundefined [1]{%
 \@ifx{#1\undefined}
}%
\providecommand \@ifnum [1]{%
 \ifnum #1\expandafter \@firstoftwo
 \else \expandafter \@secondoftwo
 \fi
}%
\providecommand \@ifx [1]{%
 \ifx #1\expandafter \@firstoftwo
 \else \expandafter \@secondoftwo
 \fi
}%
\providecommand \natexlab [1]{#1}%
\providecommand \enquote  [1]{``#1''}%
\providecommand \bibnamefont  [1]{#1}%
\providecommand \bibfnamefont [1]{#1}%
\providecommand \citenamefont [1]{#1}%
\providecommand \href@noop [0]{\@secondoftwo}%
\providecommand \href [0]{\begingroup \@sanitize@url \@href}%
\providecommand \@href[1]{\@@startlink{#1}\@@href}%
\providecommand \@@href[1]{\endgroup#1\@@endlink}%
\providecommand \@sanitize@url [0]{\catcode `\\12\catcode `\$12\catcode
  `\&12\catcode `\#12\catcode `\^12\catcode `\_12\catcode `\%12\relax}%
\providecommand \@@startlink[1]{}%
\providecommand \@@endlink[0]{}%
\providecommand \url  [0]{\begingroup\@sanitize@url \@url }%
\providecommand \@url [1]{\endgroup\@href {#1}{\urlprefix }}%
\providecommand \urlprefix  [0]{URL }%
\providecommand \Eprint [0]{\href }%
\providecommand \doibase [0]{http://dx.doi.org/}%
\providecommand \selectlanguage [0]{\@gobble}%
\providecommand \bibinfo  [0]{\@secondoftwo}%
\providecommand \bibfield  [0]{\@secondoftwo}%
\providecommand \translation [1]{[#1]}%
\providecommand \BibitemOpen [0]{}%
\providecommand \bibitemStop [0]{}%
\providecommand \bibitemNoStop [0]{.\EOS\space}%
\providecommand \EOS [0]{\spacefactor3000\relax}%
\providecommand \BibitemShut  [1]{\csname bibitem#1\endcsname}%
\let\auto@bib@innerbib\@empty
\bibitem [{\citenamefont {Wright}\ \emph {et~al.}(2000)\citenamefont {Wright},
  \citenamefont {Arlt},\ and\ \citenamefont {Dholakia}}]{WrightToroidal}%
  \BibitemOpen
  \bibfield  {author} {\bibinfo {author} {\bibfnamefont {E.~M.}\ \bibnamefont
  {Wright}}, \bibinfo {author} {\bibfnamefont {J.}~\bibnamefont {Arlt}}, \ and\
  \bibinfo {author} {\bibfnamefont {K.}~\bibnamefont {Dholakia}},\ }\bibfield
  {title} {\enquote {\bibinfo {title} {{Toroidal optical dipole traps for
  atomic Bose-Einstein condensates using Laguerre-Gaussian beams}},}\ }\href
  {\doibase 10.1103/PhysRevA.63.013608} {\bibfield  {journal} {\bibinfo
  {journal} {Phys. Rev. A}\ }\textbf {\bibinfo {volume} {63}},\ \bibinfo
  {pages} {013608} (\bibinfo {year} {2000})}\BibitemShut {NoStop}%
\bibitem [{\citenamefont {Amico}\ \emph {et~al.}(2005)\citenamefont {Amico},
  \citenamefont {Osterloh},\ and\ \citenamefont {Cataliotti}}]{Amico2005}%
  \BibitemOpen
  \bibfield  {author} {\bibinfo {author} {\bibfnamefont {Luigi}\ \bibnamefont
  {Amico}}, \bibinfo {author} {\bibfnamefont {Andreas}\ \bibnamefont
  {Osterloh}}, \ and\ \bibinfo {author} {\bibfnamefont {Francesco}\
  \bibnamefont {Cataliotti}},\ }\bibfield  {title} {\enquote {\bibinfo {title}
  {{Quantum Many Particle Systems in Ring-Shaped Optical Lattices}},}\ }\href
  {\doibase 10.1103/PhysRevLett.95.063201} {\bibfield  {journal} {\bibinfo
  {journal} {Phys. Rev. Lett.}\ }\textbf {\bibinfo {volume} {95}},\ \bibinfo
  {pages} {063201} (\bibinfo {year} {2005})}\BibitemShut {NoStop}%
\bibitem [{\citenamefont {Ryu}\ \emph {et~al.}(2007)\citenamefont {Ryu},
  \citenamefont {Andersen}, \citenamefont {Clad\'e}, \citenamefont {Natarajan},
  \citenamefont {Helmerson},\ and\ \citenamefont {Phillips}}]{RyuToroidal}%
  \BibitemOpen
  \bibfield  {author} {\bibinfo {author} {\bibfnamefont {C.}~\bibnamefont
  {Ryu}}, \bibinfo {author} {\bibfnamefont {M.~F.}\ \bibnamefont {Andersen}},
  \bibinfo {author} {\bibfnamefont {P.}~\bibnamefont {Clad\'e}}, \bibinfo
  {author} {\bibfnamefont {Vasant}\ \bibnamefont {Natarajan}}, \bibinfo
  {author} {\bibfnamefont {K.}~\bibnamefont {Helmerson}}, \ and\ \bibinfo
  {author} {\bibfnamefont {W.~D.}\ \bibnamefont {Phillips}},\ }\bibfield
  {title} {\enquote {\bibinfo {title} {{Observation of Persistent Flow of a
  Bose-Einstein Condensate in a Toroidal Trap}},}\ }\href {\doibase
  10.1103/PhysRevLett.99.260401} {\bibfield  {journal} {\bibinfo  {journal}
  {Phys. Rev. Lett.}\ }\textbf {\bibinfo {volume} {99}},\ \bibinfo {pages}
  {260401} (\bibinfo {year} {2007})}\BibitemShut {NoStop}%
\bibitem [{\citenamefont {Ramanathan}\ \emph {et~al.}(2011)\citenamefont
  {Ramanathan}, \citenamefont {Wright}, \citenamefont {Muniz}, \citenamefont
  {Zelan}, \citenamefont {Hill}, \citenamefont {Lobb}, \citenamefont
  {Helmerson}, \citenamefont {Phillips},\ and\ \citenamefont
  {Campbell}}]{RamanathanToroidal}%
  \BibitemOpen
  \bibfield  {author} {\bibinfo {author} {\bibfnamefont {A.}~\bibnamefont
  {Ramanathan}}, \bibinfo {author} {\bibfnamefont {K.~C.}\ \bibnamefont
  {Wright}}, \bibinfo {author} {\bibfnamefont {S.~R.}\ \bibnamefont {Muniz}},
  \bibinfo {author} {\bibfnamefont {M.}~\bibnamefont {Zelan}}, \bibinfo
  {author} {\bibfnamefont {W.~T.}\ \bibnamefont {Hill}}, \bibinfo {author}
  {\bibfnamefont {C.~J.}\ \bibnamefont {Lobb}}, \bibinfo {author}
  {\bibfnamefont {K.}~\bibnamefont {Helmerson}}, \bibinfo {author}
  {\bibfnamefont {W.~D.}\ \bibnamefont {Phillips}}, \ and\ \bibinfo {author}
  {\bibfnamefont {G.~K.}\ \bibnamefont {Campbell}},\ }\bibfield  {title}
  {\enquote {\bibinfo {title} {{Superflow in a Toroidal Bose-Einstein
  Condensate: An Atom Circuit with a Tunable Weak Link}},}\ }\href {\doibase
  10.1103/PhysRevLett.106.130401} {\bibfield  {journal} {\bibinfo  {journal}
  {Phys. Rev. Lett.}\ }\textbf {\bibinfo {volume} {106}},\ \bibinfo {pages}
  {130401} (\bibinfo {year} {2011})}\BibitemShut {NoStop}%
\bibitem [{\citenamefont {Corman}\ \emph {et~al.}(2014)\citenamefont {Corman},
  \citenamefont {Chomaz}, \citenamefont {Bienaim\'e}, \citenamefont
  {Desbuquois}, \citenamefont {Weitenberg}, \citenamefont {Nascimb\`ene},
  \citenamefont {Dalibard},\ and\ \citenamefont {Beugnon}}]{Corman2014}%
  \BibitemOpen
  \bibfield  {author} {\bibinfo {author} {\bibfnamefont {L.}~\bibnamefont
  {Corman}}, \bibinfo {author} {\bibfnamefont {L.}~\bibnamefont {Chomaz}},
  \bibinfo {author} {\bibfnamefont {T.}~\bibnamefont {Bienaim\'e}}, \bibinfo
  {author} {\bibfnamefont {R.}~\bibnamefont {Desbuquois}}, \bibinfo {author}
  {\bibfnamefont {C.}~\bibnamefont {Weitenberg}}, \bibinfo {author}
  {\bibfnamefont {S.}~\bibnamefont {Nascimb\`ene}}, \bibinfo {author}
  {\bibfnamefont {J.}~\bibnamefont {Dalibard}}, \ and\ \bibinfo {author}
  {\bibfnamefont {J.}~\bibnamefont {Beugnon}},\ }\bibfield  {title} {\enquote
  {\bibinfo {title} {{Quench-Induced Supercurrents in an Annular Bose Gas}},}\
  }\href {\doibase 10.1103/PhysRevLett.113.135302} {\bibfield  {journal}
  {\bibinfo  {journal} {Phys. Rev. Lett.}\ }\textbf {\bibinfo {volume} {113}},\
  \bibinfo {pages} {135302} (\bibinfo {year} {2014})}\BibitemShut {NoStop}%
\bibitem [{\citenamefont {Seaman}\ \emph {et~al.}(2007)\citenamefont {Seaman},
  \citenamefont {Kr\"amer}, \citenamefont {Anderson},\ and\ \citenamefont
  {Holland}}]{AtomtronicsFirstPaper}%
  \BibitemOpen
  \bibfield  {author} {\bibinfo {author} {\bibfnamefont {B.~T.}\ \bibnamefont
  {Seaman}}, \bibinfo {author} {\bibfnamefont {M.}~\bibnamefont {Kr\"amer}},
  \bibinfo {author} {\bibfnamefont {D.~Z.}\ \bibnamefont {Anderson}}, \ and\
  \bibinfo {author} {\bibfnamefont {M.~J.}\ \bibnamefont {Holland}},\
  }\bibfield  {title} {\enquote {\bibinfo {title} {Atomtronics: Ultracold-atom
  analogs of electronic devices},}\ }\href {\doibase
  10.1103/PhysRevA.75.023615} {\bibfield  {journal} {\bibinfo  {journal} {Phys.
  Rev. A}\ }\textbf {\bibinfo {volume} {75}},\ \bibinfo {pages} {023615}
  (\bibinfo {year} {2007})}\BibitemShut {NoStop}%
\bibitem [{\citenamefont {Amico}\ \emph {et~al.}(2014)\citenamefont {Amico},
  \citenamefont {Aghamalyan}, \citenamefont {Auksztol}, \citenamefont {Crepaz},
  \citenamefont {Dumke},\ and\ \citenamefont {Kwek}}]{Amico2014}%
  \BibitemOpen
  \bibfield  {author} {\bibinfo {author} {\bibfnamefont {Luigi}\ \bibnamefont
  {Amico}}, \bibinfo {author} {\bibfnamefont {Davit}\ \bibnamefont
  {Aghamalyan}}, \bibinfo {author} {\bibfnamefont {Filip}\ \bibnamefont
  {Auksztol}}, \bibinfo {author} {\bibfnamefont {Herbert}\ \bibnamefont
  {Crepaz}}, \bibinfo {author} {\bibfnamefont {Rainer}\ \bibnamefont {Dumke}},
  \ and\ \bibinfo {author} {\bibfnamefont {Leong~Chuan}\ \bibnamefont {Kwek}},\
  }\bibfield  {title} {\enquote {\bibinfo {title} {Superfluid qubit systems
  with ring shaped optical lattices},}\ }\href {\doibase 10.1038/srep04298}
  {\bibfield  {journal} {\bibinfo  {journal} {Scientific Reports}\ }\textbf
  {\bibinfo {volume} {4}},\ \bibinfo {pages} {4298} (\bibinfo {year}
  {2014})}\BibitemShut {NoStop}%
\bibitem [{\citenamefont {Amico}\ \emph {et~al.}(2017)\citenamefont {Amico},
  \citenamefont {Birkl}, \citenamefont {Boshier},\ and\ \citenamefont
  {Kwek}}]{Amico2017}%
  \BibitemOpen
  \bibfield  {author} {\bibinfo {author} {\bibfnamefont {Luigi}\ \bibnamefont
  {Amico}}, \bibinfo {author} {\bibfnamefont {Gerhard}\ \bibnamefont {Birkl}},
  \bibinfo {author} {\bibfnamefont {Malcolm}\ \bibnamefont {Boshier}}, \ and\
  \bibinfo {author} {\bibfnamefont {Leong-Chuan}\ \bibnamefont {Kwek}},\
  }\bibfield  {title} {\enquote {\bibinfo {title} {Focus on atomtronics-enabled
  quantum technologies},}\ }\href {\doibase 10.1088/1367-2630/aa5a6d}
  {\bibfield  {journal} {\bibinfo  {journal} {New Journal of Physics}\ }\textbf
  {\bibinfo {volume} {19}},\ \bibinfo {pages} {020201} (\bibinfo {year}
  {2017})}\BibitemShut {NoStop}%
\bibitem [{\citenamefont {Amico}\ \emph {et~al.}(2021)\citenamefont {Amico},
  \citenamefont {Boshier}, \citenamefont {Birkl}, \citenamefont {Minguzzi},
  \citenamefont {Miniatura}, \citenamefont {Kwek}, \citenamefont {Aghamalyan},
  \citenamefont {Ahufinger}, \citenamefont {Anderson}, \citenamefont {Andrei},
  \citenamefont {Arnold}, \citenamefont {Baker}, \citenamefont {Bell},
  \citenamefont {Bland}, \citenamefont {Brantut}, \citenamefont {Cassettari},
  \citenamefont {Chetcuti}, \citenamefont {Chevy}, \citenamefont {Citro},
  \citenamefont {De~Palo}, \citenamefont {Dumke}, \citenamefont {Edwards},
  \citenamefont {Folman}, \citenamefont {Fortagh}, \citenamefont {Gardiner},
  \citenamefont {Garraway}, \citenamefont {Gauthier}, \citenamefont {Günther},
  \citenamefont {Haug}, \citenamefont {Hufnagel}, \citenamefont {Keil},
  \citenamefont {Ireland}, \citenamefont {Lebrat}, \citenamefont {Li},
  \citenamefont {Longchambon}, \citenamefont {Mompart}, \citenamefont {Morsch},
  \citenamefont {Naldesi}, \citenamefont {Neely}, \citenamefont {Olshanii},
  \citenamefont {Orignac}, \citenamefont {Pandey}, \citenamefont
  {Pérez-Obiol}, \citenamefont {Perrin}, \citenamefont {Piroli}, \citenamefont
  {Polo}, \citenamefont {Pritchard}, \citenamefont {Proukakis}, \citenamefont
  {Rylands}, \citenamefont {Rubinsztein-Dunlop}, \citenamefont {Scazza},
  \citenamefont {Stringari}, \citenamefont {Tosto}, \citenamefont
  {Trombettoni}, \citenamefont {Victorin}, \citenamefont {Klitzing},
  \citenamefont {Wilkowski}, \citenamefont {Xhani},\ and\ \citenamefont
  {Yakimenko}}]{Amico2021}%
  \BibitemOpen
  \bibfield  {author} {\bibinfo {author} {\bibfnamefont {L.}~\bibnamefont
  {Amico}}, \bibinfo {author} {\bibfnamefont {M.}~\bibnamefont {Boshier}},
  \bibinfo {author} {\bibfnamefont {G.}~\bibnamefont {Birkl}}, \bibinfo
  {author} {\bibfnamefont {A.}~\bibnamefont {Minguzzi}}, \bibinfo {author}
  {\bibfnamefont {C.}~\bibnamefont {Miniatura}}, \bibinfo {author}
  {\bibfnamefont {L.-C.}\ \bibnamefont {Kwek}}, \bibinfo {author}
  {\bibfnamefont {D.}~\bibnamefont {Aghamalyan}}, \bibinfo {author}
  {\bibfnamefont {V.}~\bibnamefont {Ahufinger}}, \bibinfo {author}
  {\bibfnamefont {D.}~\bibnamefont {Anderson}}, \bibinfo {author}
  {\bibfnamefont {N.}~\bibnamefont {Andrei}}, \bibinfo {author} {\bibfnamefont
  {A.~S.}\ \bibnamefont {Arnold}}, \bibinfo {author} {\bibfnamefont
  {M.}~\bibnamefont {Baker}}, \bibinfo {author} {\bibfnamefont {T.~A.}\
  \bibnamefont {Bell}}, \bibinfo {author} {\bibfnamefont {T.}~\bibnamefont
  {Bland}}, \bibinfo {author} {\bibfnamefont {J.~P.}\ \bibnamefont {Brantut}},
  \bibinfo {author} {\bibfnamefont {D.}~\bibnamefont {Cassettari}}, \bibinfo
  {author} {\bibfnamefont {W.~J.}\ \bibnamefont {Chetcuti}}, \bibinfo {author}
  {\bibfnamefont {F.}~\bibnamefont {Chevy}}, \bibinfo {author} {\bibfnamefont
  {R.}~\bibnamefont {Citro}}, \bibinfo {author} {\bibfnamefont
  {S.}~\bibnamefont {De~Palo}}, \bibinfo {author} {\bibfnamefont
  {R.}~\bibnamefont {Dumke}}, \bibinfo {author} {\bibfnamefont
  {M.}~\bibnamefont {Edwards}}, \bibinfo {author} {\bibfnamefont
  {R.}~\bibnamefont {Folman}}, \bibinfo {author} {\bibfnamefont
  {J.}~\bibnamefont {Fortagh}}, \bibinfo {author} {\bibfnamefont {S.~A.}\
  \bibnamefont {Gardiner}}, \bibinfo {author} {\bibfnamefont {B.~M.}\
  \bibnamefont {Garraway}}, \bibinfo {author} {\bibfnamefont {G.}~\bibnamefont
  {Gauthier}}, \bibinfo {author} {\bibfnamefont {A.}~\bibnamefont {Günther}},
  \bibinfo {author} {\bibfnamefont {T.}~\bibnamefont {Haug}}, \bibinfo {author}
  {\bibfnamefont {C.}~\bibnamefont {Hufnagel}}, \bibinfo {author}
  {\bibfnamefont {M.}~\bibnamefont {Keil}}, \bibinfo {author} {\bibfnamefont
  {P.}~\bibnamefont {Ireland}}, \bibinfo {author} {\bibfnamefont
  {M.}~\bibnamefont {Lebrat}}, \bibinfo {author} {\bibfnamefont
  {W.}~\bibnamefont {Li}}, \bibinfo {author} {\bibfnamefont {L.}~\bibnamefont
  {Longchambon}}, \bibinfo {author} {\bibfnamefont {J.}~\bibnamefont
  {Mompart}}, \bibinfo {author} {\bibfnamefont {O.}~\bibnamefont {Morsch}},
  \bibinfo {author} {\bibfnamefont {P.}~\bibnamefont {Naldesi}}, \bibinfo
  {author} {\bibfnamefont {T.~W.}\ \bibnamefont {Neely}}, \bibinfo {author}
  {\bibfnamefont {M.}~\bibnamefont {Olshanii}}, \bibinfo {author}
  {\bibfnamefont {E.}~\bibnamefont {Orignac}}, \bibinfo {author} {\bibfnamefont
  {S.}~\bibnamefont {Pandey}}, \bibinfo {author} {\bibfnamefont
  {A.}~\bibnamefont {Pérez-Obiol}}, \bibinfo {author} {\bibfnamefont
  {H.}~\bibnamefont {Perrin}}, \bibinfo {author} {\bibfnamefont
  {L.}~\bibnamefont {Piroli}}, \bibinfo {author} {\bibfnamefont
  {J.}~\bibnamefont {Polo}}, \bibinfo {author} {\bibfnamefont {A.~L.}\
  \bibnamefont {Pritchard}}, \bibinfo {author} {\bibfnamefont {N.~P.}\
  \bibnamefont {Proukakis}}, \bibinfo {author} {\bibfnamefont {C.}~\bibnamefont
  {Rylands}}, \bibinfo {author} {\bibfnamefont {H.}~\bibnamefont
  {Rubinsztein-Dunlop}}, \bibinfo {author} {\bibfnamefont {F.}~\bibnamefont
  {Scazza}}, \bibinfo {author} {\bibfnamefont {S.}~\bibnamefont {Stringari}},
  \bibinfo {author} {\bibfnamefont {F.}~\bibnamefont {Tosto}}, \bibinfo
  {author} {\bibfnamefont {A.}~\bibnamefont {Trombettoni}}, \bibinfo {author}
  {\bibfnamefont {N.}~\bibnamefont {Victorin}}, \bibinfo {author}
  {\bibfnamefont {W.~von}\ \bibnamefont {Klitzing}}, \bibinfo {author}
  {\bibfnamefont {D.}~\bibnamefont {Wilkowski}}, \bibinfo {author}
  {\bibfnamefont {K.}~\bibnamefont {Xhani}}, \ and\ \bibinfo {author}
  {\bibfnamefont {A.}~\bibnamefont {Yakimenko}},\ }\bibfield  {title} {\enquote
  {\bibinfo {title} {{Roadmap on Atomtronics: State of the art and
  perspective}},}\ }\href {\doibase 10.1116/5.0026178} {\bibfield  {journal}
  {\bibinfo  {journal} {AVS Quantum Science}\ }\textbf {\bibinfo {volume}
  {3}},\ \bibinfo {pages} {039201} (\bibinfo {year} {2021})}\BibitemShut
  {NoStop}%
\bibitem [{\citenamefont {Micheli}\ \emph {et~al.}(2004)\citenamefont
  {Micheli}, \citenamefont {Daley}, \citenamefont {Jaksch},\ and\ \citenamefont
  {Zoller}}]{Micheli2004}%
  \BibitemOpen
  \bibfield  {author} {\bibinfo {author} {\bibfnamefont {A.}~\bibnamefont
  {Micheli}}, \bibinfo {author} {\bibfnamefont {A.~J.}\ \bibnamefont {Daley}},
  \bibinfo {author} {\bibfnamefont {D.}~\bibnamefont {Jaksch}}, \ and\ \bibinfo
  {author} {\bibfnamefont {P.}~\bibnamefont {Zoller}},\ }\bibfield  {title}
  {\enquote {\bibinfo {title} {{Single Atom Transistor in a 1D Optical
  Lattice}},}\ }\href {\doibase 10.1103/PhysRevLett.93.140408} {\bibfield
  {journal} {\bibinfo  {journal} {Phys. Rev. Lett.}\ }\textbf {\bibinfo
  {volume} {93}},\ \bibinfo {pages} {140408} (\bibinfo {year}
  {2004})}\BibitemShut {NoStop}%
\bibitem [{\citenamefont {Stickney}\ \emph {et~al.}(2007)\citenamefont
  {Stickney}, \citenamefont {Anderson},\ and\ \citenamefont
  {Zozulya}}]{Stickney2007}%
  \BibitemOpen
  \bibfield  {author} {\bibinfo {author} {\bibfnamefont {James~A.}\
  \bibnamefont {Stickney}}, \bibinfo {author} {\bibfnamefont {Dana~Z.}\
  \bibnamefont {Anderson}}, \ and\ \bibinfo {author} {\bibfnamefont {Alex~A.}\
  \bibnamefont {Zozulya}},\ }\bibfield  {title} {\enquote {\bibinfo {title}
  {{Transistorlike behavior of a Bose-Einstein condensate in a triple-well
  potential}},}\ }\href {\doibase 10.1103/PhysRevA.75.013608} {\bibfield
  {journal} {\bibinfo  {journal} {Phys. Rev. A}\ }\textbf {\bibinfo {volume}
  {75}},\ \bibinfo {pages} {013608} (\bibinfo {year} {2007})}\BibitemShut
  {NoStop}%
\bibitem [{\citenamefont {Ruschhaupt}\ and\ \citenamefont
  {Muga}(2007)}]{Ruschhaupt2007}%
  \BibitemOpen
  \bibfield  {author} {\bibinfo {author} {\bibfnamefont {A.}~\bibnamefont
  {Ruschhaupt}}\ and\ \bibinfo {author} {\bibfnamefont {J.~G.}\ \bibnamefont
  {Muga}},\ }\bibfield  {title} {\enquote {\bibinfo {title} {{Three-dimensional
  effects in atom diodes: Atom-optical devices for one-way motion}},}\ }\href
  {\doibase 10.1103/PhysRevA.76.013619} {\bibfield  {journal} {\bibinfo
  {journal} {Phys. Rev. A}\ }\textbf {\bibinfo {volume} {76}},\ \bibinfo
  {pages} {013619} (\bibinfo {year} {2007})}\BibitemShut {NoStop}%
\bibitem [{\citenamefont {Thorn}\ \emph {et~al.}(2008)\citenamefont {Thorn},
  \citenamefont {Schoene}, \citenamefont {Li},\ and\ \citenamefont
  {Steck}}]{Thorn2008}%
  \BibitemOpen
  \bibfield  {author} {\bibinfo {author} {\bibfnamefont {Jeremy~J.}\
  \bibnamefont {Thorn}}, \bibinfo {author} {\bibfnamefont {Elizabeth~A.}\
  \bibnamefont {Schoene}}, \bibinfo {author} {\bibfnamefont {Tao}\ \bibnamefont
  {Li}}, \ and\ \bibinfo {author} {\bibfnamefont {Daniel~A.}\ \bibnamefont
  {Steck}},\ }\bibfield  {title} {\enquote {\bibinfo {title} {Experimental
  realization of an optical one-way barrier for neutral atoms},}\ }\href
  {\doibase 10.1103/PhysRevLett.100.240407} {\bibfield  {journal} {\bibinfo
  {journal} {Phys. Rev. Lett.}\ }\textbf {\bibinfo {volume} {100}},\ \bibinfo
  {pages} {240407} (\bibinfo {year} {2008})}\BibitemShut {NoStop}%
\bibitem [{\citenamefont {Pepino}\ \emph {et~al.}(2009)\citenamefont {Pepino},
  \citenamefont {Cooper}, \citenamefont {Anderson},\ and\ \citenamefont
  {Holland}}]{Pepino2009}%
  \BibitemOpen
  \bibfield  {author} {\bibinfo {author} {\bibfnamefont {R.~A.}\ \bibnamefont
  {Pepino}}, \bibinfo {author} {\bibfnamefont {J.}~\bibnamefont {Cooper}},
  \bibinfo {author} {\bibfnamefont {D.~Z.}\ \bibnamefont {Anderson}}, \ and\
  \bibinfo {author} {\bibfnamefont {M.~J.}\ \bibnamefont {Holland}},\
  }\bibfield  {title} {\enquote {\bibinfo {title} {{Atomtronic Circuits of
  Diodes and Transistors}},}\ }\href {\doibase 10.1103/PhysRevLett.103.140405}
  {\bibfield  {journal} {\bibinfo  {journal} {Phys. Rev. Lett.}\ }\textbf
  {\bibinfo {volume} {103}},\ \bibinfo {pages} {140405} (\bibinfo {year}
  {2009})}\BibitemShut {NoStop}%
\bibitem [{\citenamefont {Krinner}\ \emph {et~al.}(2017)\citenamefont
  {Krinner}, \citenamefont {Esslinger},\ and\ \citenamefont
  {Brantut}}]{Krinner2017}%
  \BibitemOpen
  \bibfield  {author} {\bibinfo {author} {\bibfnamefont {Sebastian}\
  \bibnamefont {Krinner}}, \bibinfo {author} {\bibfnamefont {Tilman}\
  \bibnamefont {Esslinger}}, \ and\ \bibinfo {author} {\bibfnamefont
  {Jean-Philippe}\ \bibnamefont {Brantut}},\ }\bibfield  {title} {\enquote
  {\bibinfo {title} {Two-terminal transport measurements with cold atoms},}\
  }\href {\doibase 10.1088/1361-648x/aa74a1} {\bibfield  {journal} {\bibinfo
  {journal} {Journal of Physics: Condensed Matter}\ }\textbf {\bibinfo {volume}
  {29}},\ \bibinfo {pages} {343003} (\bibinfo {year} {2017})}\BibitemShut
  {NoStop}%
\bibitem [{\citenamefont {Zozulya}\ and\ \citenamefont
  {Anderson}(2013)}]{Zozulya2013}%
  \BibitemOpen
  \bibfield  {author} {\bibinfo {author} {\bibfnamefont {Alex~A.}\ \bibnamefont
  {Zozulya}}\ and\ \bibinfo {author} {\bibfnamefont {Dana~Z.}\ \bibnamefont
  {Anderson}},\ }\bibfield  {title} {\enquote {\bibinfo {title} {Principles of
  an atomtronic battery},}\ }\href {\doibase 10.1103/PhysRevA.88.043641}
  {\bibfield  {journal} {\bibinfo  {journal} {Phys. Rev. A}\ }\textbf {\bibinfo
  {volume} {88}},\ \bibinfo {pages} {043641} (\bibinfo {year}
  {2013})}\BibitemShut {NoStop}%
\bibitem [{\citenamefont {Caliga}\ \emph {et~al.}(2017)\citenamefont {Caliga},
  \citenamefont {Straatsma},\ and\ \citenamefont {Anderson}}]{Caliga2017}%
  \BibitemOpen
  \bibfield  {author} {\bibinfo {author} {\bibfnamefont {Seth~C}\ \bibnamefont
  {Caliga}}, \bibinfo {author} {\bibfnamefont {Cameron J~E}\ \bibnamefont
  {Straatsma}}, \ and\ \bibinfo {author} {\bibfnamefont {Dana~Z}\ \bibnamefont
  {Anderson}},\ }\bibfield  {title} {\enquote {\bibinfo {title} {Experimental
  demonstration of an atomtronic battery},}\ }\href {\doibase
  10.1088/1367-2630/aa56d8} {\bibfield  {journal} {\bibinfo  {journal} {New
  Journal of Physics}\ }\textbf {\bibinfo {volume} {19}},\ \bibinfo {pages}
  {013036} (\bibinfo {year} {2017})}\BibitemShut {NoStop}%
\bibitem [{\citenamefont {Caliga}\ \emph {et~al.}(2016)\citenamefont {Caliga},
  \citenamefont {Straatsma}, \citenamefont {Zozulya},\ and\ \citenamefont
  {Anderson}}]{Caliga2016}%
  \BibitemOpen
  \bibfield  {author} {\bibinfo {author} {\bibfnamefont {Seth~C}\ \bibnamefont
  {Caliga}}, \bibinfo {author} {\bibfnamefont {Cameron J~E}\ \bibnamefont
  {Straatsma}}, \bibinfo {author} {\bibfnamefont {Alex~A}\ \bibnamefont
  {Zozulya}}, \ and\ \bibinfo {author} {\bibfnamefont {Dana~Z}\ \bibnamefont
  {Anderson}},\ }\bibfield  {title} {\enquote {\bibinfo {title} {Principles of
  an atomtronic transistor},}\ }\href {\doibase 10.1088/1367-2630/18/1/015012}
  {\bibfield  {journal} {\bibinfo  {journal} {New Journal of Physics}\ }\textbf
  {\bibinfo {volume} {18}},\ \bibinfo {pages} {015012} (\bibinfo {year}
  {2016})}\BibitemShut {NoStop}%
\bibitem [{\citenamefont {Pandey}\ \emph {et~al.}(2021)\citenamefont {Pandey},
  \citenamefont {Mas}, \citenamefont {Vasilakis},\ and\ \citenamefont {von
  Klitzing}}]{Pandey2021}%
  \BibitemOpen
  \bibfield  {author} {\bibinfo {author} {\bibfnamefont {Saurabh}\ \bibnamefont
  {Pandey}}, \bibinfo {author} {\bibfnamefont {Hector}\ \bibnamefont {Mas}},
  \bibinfo {author} {\bibfnamefont {Georgios}\ \bibnamefont {Vasilakis}}, \
  and\ \bibinfo {author} {\bibfnamefont {Wolf}\ \bibnamefont {von Klitzing}},\
  }\bibfield  {title} {\enquote {\bibinfo {title} {Atomtronic matter-wave
  lensing},}\ }\href {\doibase 10.1103/PhysRevLett.126.170402} {\bibfield
  {journal} {\bibinfo  {journal} {Phys. Rev. Lett.}\ }\textbf {\bibinfo
  {volume} {126}},\ \bibinfo {pages} {170402} (\bibinfo {year}
  {2021})}\BibitemShut {NoStop}%
\bibitem [{\citenamefont {Krzyzanowska}\ \emph {et~al.}(2022)\citenamefont
  {Krzyzanowska}, \citenamefont {Ferreras}, \citenamefont {Ryu}, \citenamefont
  {Samson},\ and\ \citenamefont {Boshier}}]{Krzyzanowska2022}%
  \BibitemOpen
  \bibfield  {author} {\bibinfo {author} {\bibfnamefont {Katarzyna}\
  \bibnamefont {Krzyzanowska}}, \bibinfo {author} {\bibfnamefont {Jorge}\
  \bibnamefont {Ferreras}}, \bibinfo {author} {\bibfnamefont {Changhyun}\
  \bibnamefont {Ryu}}, \bibinfo {author} {\bibfnamefont {Edward~Carlo}\
  \bibnamefont {Samson}}, \ and\ \bibinfo {author} {\bibfnamefont {Malcolm}\
  \bibnamefont {Boshier}},\ }\href@noop {} {\enquote {\bibinfo {title} {Matter
  wave analog of a fiber-optic gyroscope},}\ } (\bibinfo {year} {2022}),\
  \Eprint {http://arxiv.org/abs/arXiv:2201.12461} {arXiv:2201.12461}
  \BibitemShut {NoStop}%
\bibitem [{\citenamefont {Wright}\ \emph {et~al.}(2013)\citenamefont {Wright},
  \citenamefont {Blakestad}, \citenamefont {Lobb}, \citenamefont {Phillips},\
  and\ \citenamefont {Campbell}}]{Wright2013}%
  \BibitemOpen
  \bibfield  {author} {\bibinfo {author} {\bibfnamefont {K.~C.}\ \bibnamefont
  {Wright}}, \bibinfo {author} {\bibfnamefont {R.~B.}\ \bibnamefont
  {Blakestad}}, \bibinfo {author} {\bibfnamefont {C.~J.}\ \bibnamefont {Lobb}},
  \bibinfo {author} {\bibfnamefont {W.~D.}\ \bibnamefont {Phillips}}, \ and\
  \bibinfo {author} {\bibfnamefont {G.~K.}\ \bibnamefont {Campbell}},\
  }\bibfield  {title} {\enquote {\bibinfo {title} {{Driving Phase Slips in a
  Superfluid Atom Circuit with a Rotating Weak Link}},}\ }\href {\doibase
  10.1103/PhysRevLett.110.025302} {\bibfield  {journal} {\bibinfo  {journal}
  {Phys. Rev. Lett.}\ }\textbf {\bibinfo {volume} {110}},\ \bibinfo {pages}
  {025302} (\bibinfo {year} {2013})}\BibitemShut {NoStop}%
\bibitem [{\citenamefont {Mathey}\ \emph {et~al.}(2014)\citenamefont {Mathey},
  \citenamefont {Clark},\ and\ \citenamefont {Mathey}}]{Mathey2014}%
  \BibitemOpen
  \bibfield  {author} {\bibinfo {author} {\bibfnamefont {Amy~C.}\ \bibnamefont
  {Mathey}}, \bibinfo {author} {\bibfnamefont {Charles~W.}\ \bibnamefont
  {Clark}}, \ and\ \bibinfo {author} {\bibfnamefont {L.}~\bibnamefont
  {Mathey}},\ }\bibfield  {title} {\enquote {\bibinfo {title} {Decay of a
  superfluid current of ultracold atoms in a toroidal trap},}\ }\href {\doibase
  10.1103/PhysRevA.90.023604} {\bibfield  {journal} {\bibinfo  {journal} {Phys.
  Rev. A}\ }\textbf {\bibinfo {volume} {90}},\ \bibinfo {pages} {023604}
  (\bibinfo {year} {2014})}\BibitemShut {NoStop}%
\bibitem [{\citenamefont {Yakimenko}\ \emph {et~al.}(2015)\citenamefont
  {Yakimenko}, \citenamefont {Isaieva}, \citenamefont {Vilchinskii},\ and\
  \citenamefont {Ostrovskaya}}]{Yakimenko2015}%
  \BibitemOpen
  \bibfield  {author} {\bibinfo {author} {\bibfnamefont {A.~I.}\ \bibnamefont
  {Yakimenko}}, \bibinfo {author} {\bibfnamefont {K.~O.}\ \bibnamefont
  {Isaieva}}, \bibinfo {author} {\bibfnamefont {S.~I.}\ \bibnamefont
  {Vilchinskii}}, \ and\ \bibinfo {author} {\bibfnamefont {E.~A.}\ \bibnamefont
  {Ostrovskaya}},\ }\bibfield  {title} {\enquote {\bibinfo {title} {{Vortex
  excitation in a stirred toroidal Bose-Einstein condensate}},}\ }\href
  {\doibase 10.1103/PhysRevA.91.023607} {\bibfield  {journal} {\bibinfo
  {journal} {Phys. Rev. A}\ }\textbf {\bibinfo {volume} {91}},\ \bibinfo
  {pages} {023607} (\bibinfo {year} {2015})}\BibitemShut {NoStop}%
\bibitem [{\citenamefont {Kumar}\ \emph {et~al.}(2016)\citenamefont {Kumar},
  \citenamefont {Anderson}, \citenamefont {Phillips}, \citenamefont {Eckel},
  \citenamefont {Campbell},\ and\ \citenamefont {Stringari}}]{Kumar2016}%
  \BibitemOpen
  \bibfield  {author} {\bibinfo {author} {\bibfnamefont {A}~\bibnamefont
  {Kumar}}, \bibinfo {author} {\bibfnamefont {N}~\bibnamefont {Anderson}},
  \bibinfo {author} {\bibfnamefont {W~D}\ \bibnamefont {Phillips}}, \bibinfo
  {author} {\bibfnamefont {S}~\bibnamefont {Eckel}}, \bibinfo {author}
  {\bibfnamefont {G~K}\ \bibnamefont {Campbell}}, \ and\ \bibinfo {author}
  {\bibfnamefont {S}~\bibnamefont {Stringari}},\ }\bibfield  {title} {\enquote
  {\bibinfo {title} {{Minimally destructive, Doppler measurement of a quantized
  flow in a ring-shaped Bose{\textendash}Einstein condensate}},}\ }\href
  {\doibase 10.1088/1367-2630/18/2/025001} {\bibfield  {journal} {\bibinfo
  {journal} {New Journal of Physics}\ }\textbf {\bibinfo {volume} {18}},\
  \bibinfo {pages} {025001} (\bibinfo {year} {2016})}\BibitemShut {NoStop}%
\bibitem [{\citenamefont {Kumar}\ \emph {et~al.}(2017)\citenamefont {Kumar},
  \citenamefont {Eckel}, \citenamefont {Jendrzejewski},\ and\ \citenamefont
  {Campbell}}]{Kumar2017}%
  \BibitemOpen
  \bibfield  {author} {\bibinfo {author} {\bibfnamefont {A.}~\bibnamefont
  {Kumar}}, \bibinfo {author} {\bibfnamefont {S.}~\bibnamefont {Eckel}},
  \bibinfo {author} {\bibfnamefont {F.}~\bibnamefont {Jendrzejewski}}, \ and\
  \bibinfo {author} {\bibfnamefont {G.~K.}\ \bibnamefont {Campbell}},\
  }\bibfield  {title} {\enquote {\bibinfo {title} {Temperature-induced decay of
  persistent currents in a superfluid ultracold gas},}\ }\href {\doibase
  10.1103/PhysRevA.95.021602} {\bibfield  {journal} {\bibinfo  {journal} {Phys.
  Rev. A}\ }\textbf {\bibinfo {volume} {95}},\ \bibinfo {pages} {021602}
  (\bibinfo {year} {2017})}\BibitemShut {NoStop}%
\bibitem [{\citenamefont {Wang}\ \emph {et~al.}(2015)\citenamefont {Wang},
  \citenamefont {Kumar}, \citenamefont {Jendrzejewski}, \citenamefont {Wilson},
  \citenamefont {Edwards}, \citenamefont {Eckel}, \citenamefont {Campbell},\
  and\ \citenamefont {Clark}}]{Wang2015}%
  \BibitemOpen
  \bibfield  {author} {\bibinfo {author} {\bibfnamefont {Yi-Hsieh}\
  \bibnamefont {Wang}}, \bibinfo {author} {\bibfnamefont {A}~\bibnamefont
  {Kumar}}, \bibinfo {author} {\bibfnamefont {F}~\bibnamefont {Jendrzejewski}},
  \bibinfo {author} {\bibfnamefont {Ryan~M}\ \bibnamefont {Wilson}}, \bibinfo
  {author} {\bibfnamefont {Mark}\ \bibnamefont {Edwards}}, \bibinfo {author}
  {\bibfnamefont {S}~\bibnamefont {Eckel}}, \bibinfo {author} {\bibfnamefont
  {G~K}\ \bibnamefont {Campbell}}, \ and\ \bibinfo {author} {\bibfnamefont
  {Charles~W}\ \bibnamefont {Clark}},\ }\bibfield  {title} {\enquote {\bibinfo
  {title} {Resonant wavepackets and shock waves in an atomtronic {SQUID}},}\
  }\href {\doibase 10.1088/1367-2630/17/12/125012} {\bibfield  {journal}
  {\bibinfo  {journal} {New Journal of Physics}\ }\textbf {\bibinfo {volume}
  {17}},\ \bibinfo {pages} {125012} (\bibinfo {year} {2015})}\BibitemShut
  {NoStop}%
\bibitem [{\citenamefont {Kunimi}\ and\ \citenamefont
  {Danshita}(2019)}]{Kunimi2019}%
  \BibitemOpen
  \bibfield  {author} {\bibinfo {author} {\bibfnamefont {Masaya}\ \bibnamefont
  {Kunimi}}\ and\ \bibinfo {author} {\bibfnamefont {Ippei}\ \bibnamefont
  {Danshita}},\ }\bibfield  {title} {\enquote {\bibinfo {title} {{Decay
  mechanisms of superflow of Bose-Einstein condensates in ring traps}},}\
  }\href {\doibase 10.1103/PhysRevA.99.043613} {\bibfield  {journal} {\bibinfo
  {journal} {Phys. Rev. A}\ }\textbf {\bibinfo {volume} {99}},\ \bibinfo
  {pages} {043613} (\bibinfo {year} {2019})}\BibitemShut {NoStop}%
\bibitem [{\citenamefont {Polo}\ \emph {et~al.}(2019)\citenamefont {Polo},
  \citenamefont {Dubessy}, \citenamefont {Pedri}, \citenamefont {Perrin},\ and\
  \citenamefont {Minguzzi}}]{Polo2019}%
  \BibitemOpen
  \bibfield  {author} {\bibinfo {author} {\bibfnamefont {Juan}\ \bibnamefont
  {Polo}}, \bibinfo {author} {\bibfnamefont {Romain}\ \bibnamefont {Dubessy}},
  \bibinfo {author} {\bibfnamefont {Paolo}\ \bibnamefont {Pedri}}, \bibinfo
  {author} {\bibfnamefont {H\'el\`ene}\ \bibnamefont {Perrin}}, \ and\ \bibinfo
  {author} {\bibfnamefont {Anna}\ \bibnamefont {Minguzzi}},\ }\bibfield
  {title} {\enquote {\bibinfo {title} {{Oscillations and Decay of Superfluid
  Currents in a One-Dimensional Bose Gas on a Ring}},}\ }\href {\doibase
  10.1103/PhysRevLett.123.195301} {\bibfield  {journal} {\bibinfo  {journal}
  {Phys. Rev. Lett.}\ }\textbf {\bibinfo {volume} {123}},\ \bibinfo {pages}
  {195301} (\bibinfo {year} {2019})}\BibitemShut {NoStop}%
\bibitem [{\citenamefont {Zhang}\ and\ \citenamefont {Li}(2019)}]{Zhang2019}%
  \BibitemOpen
  \bibfield  {author} {\bibinfo {author} {\bibfnamefont {Xiu-Rong}\
  \bibnamefont {Zhang}}\ and\ \bibinfo {author} {\bibfnamefont {Wei-Dong}\
  \bibnamefont {Li}},\ }\bibfield  {title} {\enquote {\bibinfo {title}
  {{Current{\textendash}phase relations of a ring-trapped
  Bose{\textendash}Einstein condensate with a weak link}},}\ }\href {\doibase
  10.1088/1674-1056/28/1/010303} {\bibfield  {journal} {\bibinfo  {journal}
  {Chinese Physics B}\ }\textbf {\bibinfo {volume} {28}},\ \bibinfo {pages}
  {010303} (\bibinfo {year} {2019})}\BibitemShut {NoStop}%
\bibitem [{\citenamefont {Syafwan}\ \emph {et~al.}(2016)\citenamefont
  {Syafwan}, \citenamefont {Kevrekidis}, \citenamefont {Paris-Mandoki},
  \citenamefont {Lesanovsky}, \citenamefont {Krüger}, \citenamefont
  {Hackermüller},\ and\ \citenamefont {Susanto}}]{Syafwan2016}%
  \BibitemOpen
  \bibfield  {author} {\bibinfo {author} {\bibfnamefont {M}~\bibnamefont
  {Syafwan}}, \bibinfo {author} {\bibfnamefont {P}~\bibnamefont {Kevrekidis}},
  \bibinfo {author} {\bibfnamefont {A}~\bibnamefont {Paris-Mandoki}}, \bibinfo
  {author} {\bibfnamefont {I}~\bibnamefont {Lesanovsky}}, \bibinfo {author}
  {\bibfnamefont {P}~\bibnamefont {Krüger}}, \bibinfo {author} {\bibfnamefont
  {L}~\bibnamefont {Hackermüller}}, \ and\ \bibinfo {author} {\bibfnamefont
  {H}~\bibnamefont {Susanto}},\ }\bibfield  {title} {\enquote {\bibinfo {title}
  {{Superfluid flow past an obstacle in annular Bose{\textendash}Einstein
  condensates}},}\ }\href {\doibase 10.1088/0953-4075/49/23/235301} {\bibfield
  {journal} {\bibinfo  {journal} {Journal of Physics B: Atomic, Molecular and
  Optical Physics}\ }\textbf {\bibinfo {volume} {49}},\ \bibinfo {pages}
  {235301} (\bibinfo {year} {2016})}\BibitemShut {NoStop}%
\bibitem [{\citenamefont {Bland}\ \emph {et~al.}(2020)\citenamefont {Bland},
  \citenamefont {Marolleau}, \citenamefont {Comaron}, \citenamefont {Malomed},\
  and\ \citenamefont {Proukakis}}]{Bland2020}%
  \BibitemOpen
  \bibfield  {author} {\bibinfo {author} {\bibfnamefont {T}~\bibnamefont
  {Bland}}, \bibinfo {author} {\bibfnamefont {Q}~\bibnamefont {Marolleau}},
  \bibinfo {author} {\bibfnamefont {P}~\bibnamefont {Comaron}}, \bibinfo
  {author} {\bibfnamefont {B~A}\ \bibnamefont {Malomed}}, \ and\ \bibinfo
  {author} {\bibfnamefont {N~P}\ \bibnamefont {Proukakis}},\ }\bibfield
  {title} {\enquote {\bibinfo {title} {Persistent current formation in
  double-ring geometries},}\ }\href {\doibase 10.1088/1361-6455/ab81e9}
  {\bibfield  {journal} {\bibinfo  {journal} {Journal of Physics B: Atomic,
  Molecular and Optical Physics}\ }\textbf {\bibinfo {volume} {53}},\ \bibinfo
  {pages} {115301} (\bibinfo {year} {2020})}\BibitemShut {NoStop}%
\bibitem [{\citenamefont {Mehdi}\ \emph {et~al.}(2021)\citenamefont {Mehdi},
  \citenamefont {Bradley}, \citenamefont {Hope},\ and\ \citenamefont
  {Szigeti}}]{Mehdi2021}%
  \BibitemOpen
  \bibfield  {author} {\bibinfo {author} {\bibfnamefont {Zain}\ \bibnamefont
  {Mehdi}}, \bibinfo {author} {\bibfnamefont {Ashton~S.}\ \bibnamefont
  {Bradley}}, \bibinfo {author} {\bibfnamefont {Joseph~J.}\ \bibnamefont
  {Hope}}, \ and\ \bibinfo {author} {\bibfnamefont {Stuart~S.}\ \bibnamefont
  {Szigeti}},\ }\bibfield  {title} {\enquote {\bibinfo {title} {{Superflow
  decay in a toroidal Bose gas: The effect of quantum and thermal
  fluctuations}},}\ }\href {\doibase 10.21468/SciPostPhys.11.4.080} {\bibfield
  {journal} {\bibinfo  {journal} {SciPost Phys.}\ }\textbf {\bibinfo {volume}
  {11}},\ \bibinfo {pages} {80} (\bibinfo {year} {2021})}\BibitemShut {NoStop}%
\bibitem [{\citenamefont {Clarke}\ and\ \citenamefont
  {Braginski}(2006)}]{clarke2006squid}%
  \BibitemOpen
  \bibfield  {author} {\bibinfo {author} {\bibfnamefont {John}\ \bibnamefont
  {Clarke}}\ and\ \bibinfo {author} {\bibfnamefont {Alex~I}\ \bibnamefont
  {Braginski}},\ }\href@noop {} {\emph {\bibinfo {title} {The SQUID handbook:
  Applications of SQUIDs and SQUID systems}}}\ (\bibinfo  {publisher} {John
  Wiley \& Sons},\ \bibinfo {year} {2006})\BibitemShut {NoStop}%
\bibitem [{\citenamefont {Degen}\ \emph {et~al.}(2017)\citenamefont {Degen},
  \citenamefont {Reinhard},\ and\ \citenamefont {Cappellaro}}]{Degen2017}%
  \BibitemOpen
  \bibfield  {author} {\bibinfo {author} {\bibfnamefont {C.~L.}\ \bibnamefont
  {Degen}}, \bibinfo {author} {\bibfnamefont {F.}~\bibnamefont {Reinhard}}, \
  and\ \bibinfo {author} {\bibfnamefont {P.}~\bibnamefont {Cappellaro}},\
  }\bibfield  {title} {\enquote {\bibinfo {title} {Quantum sensing},}\ }\href
  {\doibase 10.1103/RevModPhys.89.035002} {\bibfield  {journal} {\bibinfo
  {journal} {Rev. Mod. Phys.}\ }\textbf {\bibinfo {volume} {89}},\ \bibinfo
  {pages} {035002} (\bibinfo {year} {2017})}\BibitemShut {NoStop}%
\bibitem [{\citenamefont {Ladd}\ \emph {et~al.}(2010)\citenamefont {Ladd},
  \citenamefont {Jelezko}, \citenamefont {Laflamme}, \citenamefont {Nakamura},
  \citenamefont {Monroe},\ and\ \citenamefont {O'Brien}}]{Ladd2010}%
  \BibitemOpen
  \bibfield  {author} {\bibinfo {author} {\bibfnamefont {T.~D.}\ \bibnamefont
  {Ladd}}, \bibinfo {author} {\bibfnamefont {F.}~\bibnamefont {Jelezko}},
  \bibinfo {author} {\bibfnamefont {R.}~\bibnamefont {Laflamme}}, \bibinfo
  {author} {\bibfnamefont {Y.}~\bibnamefont {Nakamura}}, \bibinfo {author}
  {\bibfnamefont {C.}~\bibnamefont {Monroe}}, \ and\ \bibinfo {author}
  {\bibfnamefont {J.~L.}\ \bibnamefont {O'Brien}},\ }\bibfield  {title}
  {\enquote {\bibinfo {title} {Quantum computers},}\ }\href {\doibase
  10.1038/nature08812} {\bibfield  {journal} {\bibinfo  {journal} {Nature}\
  }\textbf {\bibinfo {volume} {464}},\ \bibinfo {pages} {45--53} (\bibinfo
  {year} {2010})}\BibitemShut {NoStop}%
\bibitem [{\citenamefont {Eckel}\ \emph {et~al.}(2014)\citenamefont {Eckel},
  \citenamefont {Lee}, \citenamefont {Jendrzejewski}, \citenamefont {Murray},
  \citenamefont {Clark}, \citenamefont {Lobb}, \citenamefont {Phillips},
  \citenamefont {Edwards},\ and\ \citenamefont {Campbell}}]{Eckel2014}%
  \BibitemOpen
  \bibfield  {author} {\bibinfo {author} {\bibfnamefont {Stephen}\ \bibnamefont
  {Eckel}}, \bibinfo {author} {\bibfnamefont {Jeffrey~G.}\ \bibnamefont {Lee}},
  \bibinfo {author} {\bibfnamefont {Fred}\ \bibnamefont {Jendrzejewski}},
  \bibinfo {author} {\bibfnamefont {Noel}\ \bibnamefont {Murray}}, \bibinfo
  {author} {\bibfnamefont {Charles~W.}\ \bibnamefont {Clark}}, \bibinfo
  {author} {\bibfnamefont {Christopher~J.}\ \bibnamefont {Lobb}}, \bibinfo
  {author} {\bibfnamefont {William~D.}\ \bibnamefont {Phillips}}, \bibinfo
  {author} {\bibfnamefont {Mark}\ \bibnamefont {Edwards}}, \ and\ \bibinfo
  {author} {\bibfnamefont {Gretchen~K.}\ \bibnamefont {Campbell}},\ }\bibfield
  {title} {\enquote {\bibinfo {title} {{Hysteresis in a quantized superfluid
  `atomtronic'circuit}},}\ }\href {\doibase 10.1038/nature12958} {\bibfield
  {journal} {\bibinfo  {journal} {Nature}\ }\textbf {\bibinfo {volume} {506}},\
  \bibinfo {pages} {200--203} (\bibinfo {year} {2014})}\BibitemShut {NoStop}%
\bibitem [{\citenamefont {Ryu}\ \emph {et~al.}(2013)\citenamefont {Ryu},
  \citenamefont {Blackburn}, \citenamefont {Blinova},\ and\ \citenamefont
  {Boshier}}]{Ryu2013}%
  \BibitemOpen
  \bibfield  {author} {\bibinfo {author} {\bibfnamefont {C.}~\bibnamefont
  {Ryu}}, \bibinfo {author} {\bibfnamefont {P.~W.}\ \bibnamefont {Blackburn}},
  \bibinfo {author} {\bibfnamefont {A.~A.}\ \bibnamefont {Blinova}}, \ and\
  \bibinfo {author} {\bibfnamefont {M.~G.}\ \bibnamefont {Boshier}},\
  }\bibfield  {title} {\enquote {\bibinfo {title} {Experimental realization of
  josephson junctions for an atom squid},}\ }\href {\doibase
  10.1103/PhysRevLett.111.205301} {\bibfield  {journal} {\bibinfo  {journal}
  {Phys. Rev. Lett.}\ }\textbf {\bibinfo {volume} {111}},\ \bibinfo {pages}
  {205301} (\bibinfo {year} {2013})}\BibitemShut {NoStop}%
\bibitem [{\citenamefont {Jendrzejewski}\ \emph {et~al.}(2014)\citenamefont
  {Jendrzejewski}, \citenamefont {Eckel}, \citenamefont {Murray}, \citenamefont
  {Lanier}, \citenamefont {Edwards}, \citenamefont {Lobb},\ and\ \citenamefont
  {Campbell}}]{Campbell2014}%
  \BibitemOpen
  \bibfield  {author} {\bibinfo {author} {\bibfnamefont {F.}~\bibnamefont
  {Jendrzejewski}}, \bibinfo {author} {\bibfnamefont {S.}~\bibnamefont
  {Eckel}}, \bibinfo {author} {\bibfnamefont {N.}~\bibnamefont {Murray}},
  \bibinfo {author} {\bibfnamefont {C.}~\bibnamefont {Lanier}}, \bibinfo
  {author} {\bibfnamefont {M.}~\bibnamefont {Edwards}}, \bibinfo {author}
  {\bibfnamefont {C.~J.}\ \bibnamefont {Lobb}}, \ and\ \bibinfo {author}
  {\bibfnamefont {G.~K.}\ \bibnamefont {Campbell}},\ }\bibfield  {title}
  {\enquote {\bibinfo {title} {{Resistive Flow in a Weakly Interacting
  Bose-Einstein Condensate}},}\ }\href {\doibase
  10.1103/PhysRevLett.113.045305} {\bibfield  {journal} {\bibinfo  {journal}
  {Phys. Rev. Lett.}\ }\textbf {\bibinfo {volume} {113}},\ \bibinfo {pages}
  {045305} (\bibinfo {year} {2014})}\BibitemShut {NoStop}%
\bibitem [{\citenamefont {Ryu}\ \emph {et~al.}(2020)\citenamefont {Ryu},
  \citenamefont {Samson},\ and\ \citenamefont {Boshier}}]{Ryu2020}%
  \BibitemOpen
  \bibfield  {author} {\bibinfo {author} {\bibfnamefont {C.}~\bibnamefont
  {Ryu}}, \bibinfo {author} {\bibfnamefont {E.~C.}\ \bibnamefont {Samson}}, \
  and\ \bibinfo {author} {\bibfnamefont {M.~G.}\ \bibnamefont {Boshier}},\
  }\bibfield  {title} {\enquote {\bibinfo {title} {Quantum interference of
  currents in an atomtronic squid},}\ }\href {\doibase
  10.1038/s41467-020-17185-6} {\bibfield  {journal} {\bibinfo  {journal}
  {Nature Communications}\ }\textbf {\bibinfo {volume} {11}},\ \bibinfo {pages}
  {3338} (\bibinfo {year} {2020})}\BibitemShut {NoStop}%
\bibitem [{\citenamefont {Mathey}\ and\ \citenamefont
  {Mathey}(2016)}]{Mathey2016}%
  \BibitemOpen
  \bibfield  {author} {\bibinfo {author} {\bibfnamefont {Amy~C}\ \bibnamefont
  {Mathey}}\ and\ \bibinfo {author} {\bibfnamefont {L}~\bibnamefont {Mathey}},\
  }\bibfield  {title} {\enquote {\bibinfo {title} {Realizing and optimizing an
  atomtronic {SQUID}},}\ }\href {\doibase 10.1088/1367-2630/18/5/055016}
  {\bibfield  {journal} {\bibinfo  {journal} {New Journal of Physics}\ }\textbf
  {\bibinfo {volume} {18}},\ \bibinfo {pages} {055016} (\bibinfo {year}
  {2016})}\BibitemShut {NoStop}%
\bibitem [{\citenamefont {Mora}\ and\ \citenamefont {Castin}(2003)}]{Mora2003}%
  \BibitemOpen
  \bibfield  {author} {\bibinfo {author} {\bibfnamefont {Christophe}\
  \bibnamefont {Mora}}\ and\ \bibinfo {author} {\bibfnamefont {Yvan}\
  \bibnamefont {Castin}},\ }\bibfield  {title} {\enquote {\bibinfo {title}
  {{Extension of Bogoliubov theory to quasicondensates}},}\ }\href {\doibase
  10.1103/PhysRevA.67.053615} {\bibfield  {journal} {\bibinfo  {journal} {Phys.
  Rev. A}\ }\textbf {\bibinfo {volume} {67}},\ \bibinfo {pages} {053615}
  (\bibinfo {year} {2003})}\BibitemShut {NoStop}%
\bibitem [{\citenamefont {Blakie}\ \emph {et~al.}(2008)\citenamefont {Blakie},
  \citenamefont {Bradley}, \citenamefont {Davis}, \citenamefont {Ballagh},\
  and\ \citenamefont {Gardiner}}]{Blakie2008}%
  \BibitemOpen
  \bibfield  {author} {\bibinfo {author} {\bibfnamefont {P.~B.}\ \bibnamefont
  {Blakie}}, \bibinfo {author} {\bibfnamefont {A.~S.}\ \bibnamefont {Bradley}},
  \bibinfo {author} {\bibfnamefont {M.~J.}\ \bibnamefont {Davis}}, \bibinfo
  {author} {\bibfnamefont {R.~J.}\ \bibnamefont {Ballagh}}, \ and\ \bibinfo
  {author} {\bibfnamefont {C.~W.}\ \bibnamefont {Gardiner}},\ }\bibfield
  {title} {\enquote {\bibinfo {title} {{Dynamics and statistical mechanics of
  ultra-cold Bose gases using c-field techniques}},}\ }\href {\doibase
  10.1080/00018730802564254} {\bibfield  {journal} {\bibinfo  {journal}
  {Advances in Physics}\ }\textbf {\bibinfo {volume} {57}},\ \bibinfo {pages}
  {363--455} (\bibinfo {year} {2008})}\BibitemShut {NoStop}%
\bibitem [{\citenamefont {Polkovnikov}(2010)}]{Polkovnikov2010}%
  \BibitemOpen
  \bibfield  {author} {\bibinfo {author} {\bibfnamefont {Anatoli}\ \bibnamefont
  {Polkovnikov}},\ }\bibfield  {title} {\enquote {\bibinfo {title} {Phase space
  representation of quantum dynamics},}\ }\href {\doibase
  https://doi.org/10.1016/j.aop.2010.02.006} {\bibfield  {journal} {\bibinfo
  {journal} {Annals of Physics}\ }\textbf {\bibinfo {volume} {325}},\ \bibinfo
  {pages} {1790--1852} (\bibinfo {year} {2010})}\BibitemShut {NoStop}%
\bibitem [{\citenamefont {Tinkham}(2004)}]{tinkham2004}%
  \BibitemOpen
  \bibfield  {author} {\bibinfo {author} {\bibfnamefont {Michael}\ \bibnamefont
  {Tinkham}},\ }\href {http://www.worldcat.org/isbn/0486435032} {\emph
  {\bibinfo {title} {Introduction to Superconductivity}}},\ \bibinfo {edition}
  {2nd}\ ed.\ (\bibinfo  {publisher} {Dover Publications},\ \bibinfo {year}
  {2004})\BibitemShut {NoStop}%
\bibitem [{\citenamefont {Singh}\ \emph {et~al.}(2016)\citenamefont {Singh},
  \citenamefont {Weimer}, \citenamefont {Morgener}, \citenamefont {Siegl},
  \citenamefont {Hueck}, \citenamefont {Luick}, \citenamefont {Moritz},\ and\
  \citenamefont {Mathey}}]{Singh2016}%
  \BibitemOpen
  \bibfield  {author} {\bibinfo {author} {\bibfnamefont {Vijay~Pal}\
  \bibnamefont {Singh}}, \bibinfo {author} {\bibfnamefont {Wolf}\ \bibnamefont
  {Weimer}}, \bibinfo {author} {\bibfnamefont {Kai}\ \bibnamefont {Morgener}},
  \bibinfo {author} {\bibfnamefont {Jonas}\ \bibnamefont {Siegl}}, \bibinfo
  {author} {\bibfnamefont {Klaus}\ \bibnamefont {Hueck}}, \bibinfo {author}
  {\bibfnamefont {Niclas}\ \bibnamefont {Luick}}, \bibinfo {author}
  {\bibfnamefont {Henning}\ \bibnamefont {Moritz}}, \ and\ \bibinfo {author}
  {\bibfnamefont {Ludwig}\ \bibnamefont {Mathey}},\ }\bibfield  {title}
  {\enquote {\bibinfo {title} {{Probing superfluidity of Bose-Einstein
  condensates via laser stirring}},}\ }\href {\doibase
  10.1103/PhysRevA.93.023634} {\bibfield  {journal} {\bibinfo  {journal} {Phys.
  Rev. A}\ }\textbf {\bibinfo {volume} {93}},\ \bibinfo {pages} {023634}
  (\bibinfo {year} {2016})}\BibitemShut {NoStop}%
\bibitem [{\citenamefont {Kiehn}\ \emph {et~al.}(2021)\citenamefont {Kiehn},
  \citenamefont {Singh},\ and\ \citenamefont {Mathey}}]{Kiehn2021}%
  \BibitemOpen
  \bibfield  {author} {\bibinfo {author} {\bibfnamefont {Hannes}\ \bibnamefont
  {Kiehn}}, \bibinfo {author} {\bibfnamefont {Vijay~Pal}\ \bibnamefont
  {Singh}}, \ and\ \bibinfo {author} {\bibfnamefont {Ludwig}\ \bibnamefont
  {Mathey}},\ }\href@noop {} {\enquote {\bibinfo {title} {{Superfluidity of a
  laser-stirred Bose-Einstein condensate}},}\ } (\bibinfo {year} {2021}),\
  \Eprint {http://arxiv.org/abs/2110.14634} {arXiv:2110.14634
  [cond-mat.quant-gas]} \BibitemShut {NoStop}%
\bibitem [{\citenamefont {Levy}\ \emph {et~al.}(2007)\citenamefont {Levy},
  \citenamefont {Lahoud}, \citenamefont {Shomroni},\ and\ \citenamefont
  {Steinhauer}}]{Levy2007}%
  \BibitemOpen
  \bibfield  {author} {\bibinfo {author} {\bibfnamefont {S.}~\bibnamefont
  {Levy}}, \bibinfo {author} {\bibfnamefont {E.}~\bibnamefont {Lahoud}},
  \bibinfo {author} {\bibfnamefont {I.}~\bibnamefont {Shomroni}}, \ and\
  \bibinfo {author} {\bibfnamefont {J.}~\bibnamefont {Steinhauer}},\ }\bibfield
   {title} {\enquote {\bibinfo {title} {{The a.c. and d.c. Josephson effects in
  a Bose--Einstein condensate}},}\ }\href {\doibase 10.1038/nature06186}
  {\bibfield  {journal} {\bibinfo  {journal} {Nature}\ }\textbf {\bibinfo
  {volume} {449}},\ \bibinfo {pages} {579--583} (\bibinfo {year}
  {2007})}\BibitemShut {NoStop}%
\bibitem [{\citenamefont {Albiez}\ \emph {et~al.}(2005)\citenamefont {Albiez},
  \citenamefont {Gati}, \citenamefont {F\"olling}, \citenamefont {Hunsmann},
  \citenamefont {Cristiani},\ and\ \citenamefont {Oberthaler}}]{Albiez2005}%
  \BibitemOpen
  \bibfield  {author} {\bibinfo {author} {\bibfnamefont {Michael}\ \bibnamefont
  {Albiez}}, \bibinfo {author} {\bibfnamefont {Rudolf}\ \bibnamefont {Gati}},
  \bibinfo {author} {\bibfnamefont {Jonas}\ \bibnamefont {F\"olling}}, \bibinfo
  {author} {\bibfnamefont {Stefan}\ \bibnamefont {Hunsmann}}, \bibinfo {author}
  {\bibfnamefont {Matteo}\ \bibnamefont {Cristiani}}, \ and\ \bibinfo {author}
  {\bibfnamefont {Markus~K.}\ \bibnamefont {Oberthaler}},\ }\bibfield  {title}
  {\enquote {\bibinfo {title} {{Direct Observation of Tunneling and Nonlinear
  Self-Trapping in a Single Bosonic Josephson Junction}},}\ }\href {\doibase
  10.1103/PhysRevLett.95.010402} {\bibfield  {journal} {\bibinfo  {journal}
  {Phys. Rev. Lett.}\ }\textbf {\bibinfo {volume} {95}},\ \bibinfo {pages}
  {010402} (\bibinfo {year} {2005})}\BibitemShut {NoStop}%
\bibitem [{\citenamefont {Pandey}\ \emph {et~al.}(2019)\citenamefont {Pandey},
  \citenamefont {Mas}, \citenamefont {Drougakis}, \citenamefont {Thekkeppatt},
  \citenamefont {Bolpasi}, \citenamefont {Vasilakis}, \citenamefont {Poulios},\
  and\ \citenamefont {von Klitzing}}]{Pandey2019}%
  \BibitemOpen
  \bibfield  {author} {\bibinfo {author} {\bibfnamefont {Saurabh}\ \bibnamefont
  {Pandey}}, \bibinfo {author} {\bibfnamefont {Hector}\ \bibnamefont {Mas}},
  \bibinfo {author} {\bibfnamefont {Giannis}\ \bibnamefont {Drougakis}},
  \bibinfo {author} {\bibfnamefont {Premjith}\ \bibnamefont {Thekkeppatt}},
  \bibinfo {author} {\bibfnamefont {Vasiliki}\ \bibnamefont {Bolpasi}},
  \bibinfo {author} {\bibfnamefont {Georgios}\ \bibnamefont {Vasilakis}},
  \bibinfo {author} {\bibfnamefont {Konstantinos}\ \bibnamefont {Poulios}}, \
  and\ \bibinfo {author} {\bibfnamefont {Wolf}\ \bibnamefont {von Klitzing}},\
  }\bibfield  {title} {\enquote {\bibinfo {title} {{Hypersonic Bose--Einstein
  condensates in accelerator rings}},}\ }\href {\doibase
  10.1038/s41586-019-1273-5} {\bibfield  {journal} {\bibinfo  {journal}
  {Nature}\ }\textbf {\bibinfo {volume} {570}},\ \bibinfo {pages} {205--209}
  (\bibinfo {year} {2019})}\BibitemShut {NoStop}%
\bibitem [{\citenamefont {Gross}\ and\ \citenamefont {Marx}()}]{GrossNotes}%
  \BibitemOpen
  \bibfield  {author} {\bibinfo {author} {\bibfnamefont {Rudolf}\ \bibnamefont
  {Gross}}\ and\ \bibinfo {author} {\bibfnamefont {Achim}\ \bibnamefont
  {Marx}},\ }\href
  {https://www.wmi.badw.de/fileadmin/WMI/Lecturenotes/Applied_Superconductivity/AS_Chapter4.pdf}
  {\enquote {\bibinfo {title} {{Lecture notes on applied superconductivity}},}\
  }\BibitemShut {NoStop}%
\end{thebibliography}%

\end{document}